\documentclass
[aps,prb,twocolumn,floatfix,english,showpacs,10pt]{revtex4-2}%
\usepackage{graphicx}
\usepackage{booktabs}
\usepackage{amsmath}
\usepackage{physics}
\usepackage{amssymb}
\usepackage{colordvi}
\usepackage{verbatim}
\usepackage{xcolor}
\usepackage{mathrsfs}
\usepackage{epsfig}
\usepackage{lipsum}
\usepackage{amsfonts}
\usepackage{makecell}
\usepackage{esint}
\usepackage[unicode=true, breaklinks=false, pdfborder={0 0 1}, backref=false,
colorlinks=true, linkcolor=blue, urlcolor=blue, citecolor=blue]{hyperref}%
\providecommand{\U}[1]{\protect\rule{.1in}{.1in}}
\providecommand{\U}[1]{\protect\rule{.1in}{.1in}}
\setcitestyle{numbers,square}

\begin{document}

\title{Magnon correlation enables spin injection, dephasing, and transport in canted antiferromagnets}

\author{Xiyin Ye}
\affiliation{School of Physics, Huazhong University of Science and Technology, Wuhan 430074, China}

\author{Tao Yu}
\email{taoyuphy@hust.edu.cn}
\affiliation{School of Physics, Huazhong University of Science and Technology, Wuhan 430074, China}

\date{\today }

\begin{abstract}
Thermal and electrical injection and transport of magnon spins in magnetic insulators is conventionally understood by the non-equilibrium population of magnons. However, this view is challenged by several recent experiments in noncollinear antiferromagnets, which urge a thorough theoretical investigation at the fundamental level. We find that the magnon spin in noncollinear magnets is described by a matrix, so even when the diagonal terms---spins carried by population---vanish, the off-diagonal correlations transmit magnon spins. Our quantum theory shows that a net spin-flip of electrons in adjacent conductors creates quantum coherence between magnon states, which transports magnon spins in canted antiferromagnets, even without a definite phase difference between magnon modes in the incoherent process. It reveals that the pumped magnon correlation is not conserved due to an intrinsic spin torque, which causes dephasing and strong spatial spin oscillations during transport; both are enhanced by magnetic fields. Spin transfer to proximity conductors can cause extrinsic dephasing, which suppresses spin oscillations and thereby gates spin transport. 

\end{abstract}

\maketitle

\section{Introduction}

Antiferromagnets (AFMs) hold significant promise for spintronic applications due to their terahertz-frequency dynamics, immunity to external magnetic disorder, and ultrafast responses~\cite{AFM1,AFM2,AFM3,AFM4,AFM5}.
Spin angular momentum in collinear easy-axis AFMs is carried by circularly polarized magnon eigenmodes, which shows low dissipation in insulators with efficient long-range spin transport~\cite{easy_axis1,easy_axis2,easy_axis3,easy_axis4,easy_axis5,easy_axis6,easy_axis7,easy_axis8,easy_axis9,easy_axis10,YFeO,easy_axis11,easy_axis12,easy_axis13,easy_axis14,easy_axis15,easy_axis16,easy_axis17,YFeO2}.
In contrast, magnon modes in easy-plane AFMs are linearly polarized and are believed to carry no spins~\cite{AFM3,easy_plane1,easy_plane2,local_DMI} and the modes in canted AFMs have no spin projection along the N\'eel vector; both appear to be less likely to support spin currents~\cite{AFM3,YFeO,easy_plane3,easy_plane4}. Experiments on canted easy-plane hematite ($\alpha$-Fe$_2$O$_3$) deny this intuition by observing efficient long-range magnon spin transport and a field-tunable Hanle effect~\cite{easy_plane5,easy_plane6,easy_plane7}. The observation is interpreted as an interference of two linearly polarized modes, occasionally mimicking a circular polarization or pseudospin~\cite{AFM5,easy_plane8,easy_plane9,easy_plane12}.

This pseudospin description so far is classical in the sense that two classical waves with a time-dependent phase difference interfere~\cite{easy_plane13}. On the other hand, the thermal and electrical spin injection via the coupling between electrons and magnons across interfaces is a highly incoherent process since no definite initial phase difference between two waves is generated~\cite{easy_axis1,easy_axis4,easy_axis9,easy_axis10,easy_axis11,YFeO,mu2,interface22,spin_transfer1,spin_transfer2,spin_transfer3}, different from microwave excitation~\cite{easy_plane13} or spin pumping from AFMs to  metals~\cite{easy_axis14,easy_axis15,easy_axis16,easy_plane10,easy_plane11,easy_plane14,interface23}. 
This essentially raises a fundamental question: How is observable macroscopic quantum coherence established between different magnon branches in the incoherent thermal/electrical injection, and how do they dephase to the environment?

\begin{figure}[htp!]	
\includegraphics[width=0.48\textwidth,trim=0.6cm 0cm 0cm 0.1cm]{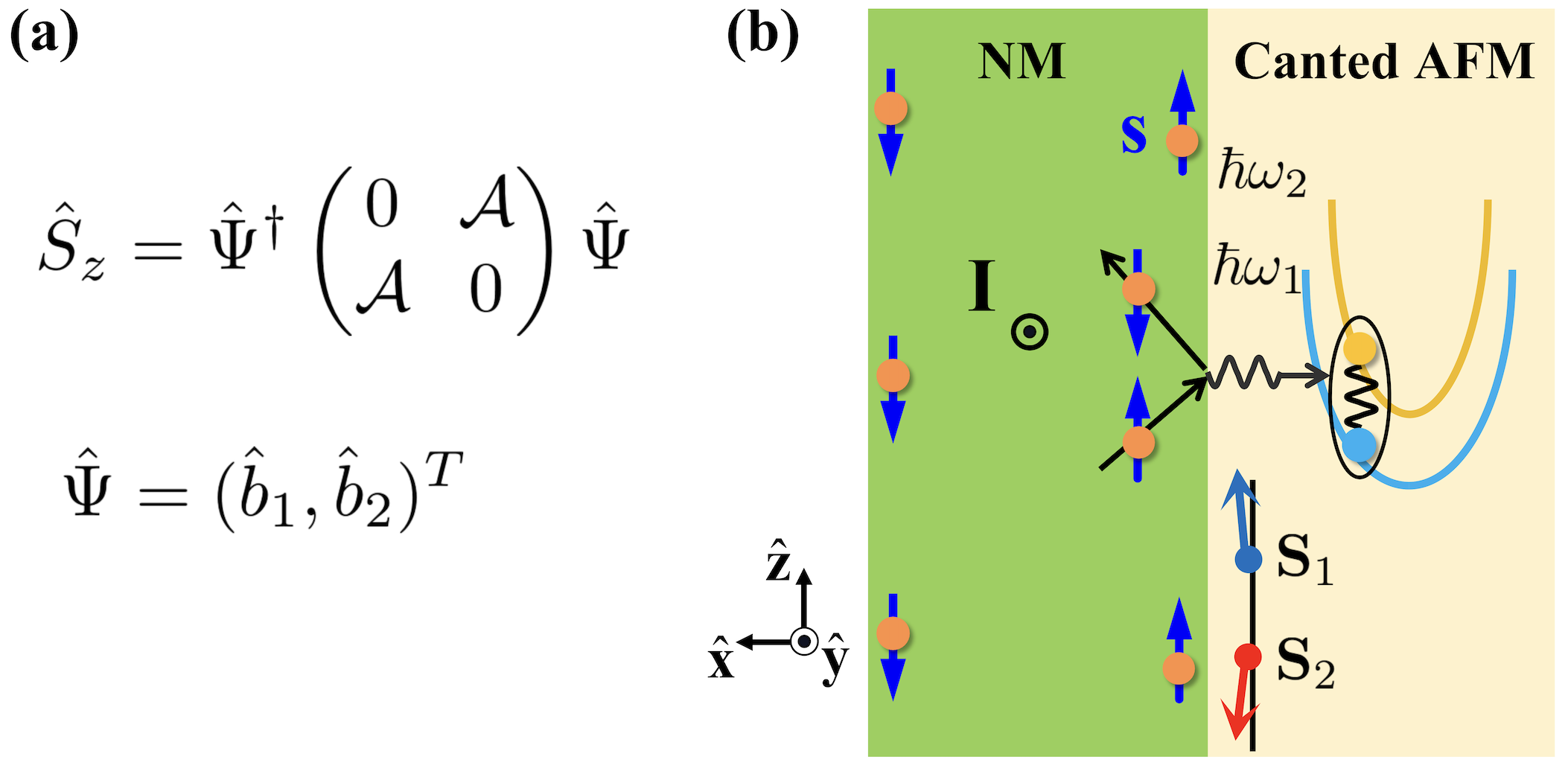}
\caption{(a) Description of magnon spin in AFMs in terms of a matrix. (b) A net spin-flip of an electron in normal metals creates the magnon correlation between two bands in AFMs. An optional configuration to pump magnon correlation is to use canted AFMs with N\'eel vector along $\hat{\bf z}$ driven by a spin accumulation $\parallel \hat{\bf z} $ created by a charge current $I$ along $\hat{\bf y}$  via the spin Hall effect.}
\label{inj_model}
\end{figure}

In this work, we demonstrate in the quantum description that the pseudospin in the classical description is a ``real spin" carried by magnons by showing that the magnon spin is described by a matrix.
When the spin carried by magnon population vanishes due to specific interactions, as shown in Fig.~\ref{inj_model}(a), the spin is solely carried by correlation ${\cal A}$, represented by the wavy line between two magnon bands $\hbar \omega_1$ and $\hbar \omega_2$ in Fig.~\ref{inj_model}(b). By developing a quantum kinetic theory for electrical spin injection, dephasing, and transport in noncollinear AFMs, we show that a net spin-flip process of electrons creates definite magnon correlations, illustrated in Fig.~\ref{inj_model}(b) with canted AFMs as a prototypical example. 
The injected magnon correlation experiences dephasing due to free-induction decay, governed by the inhomogeneous broadening~\cite{broadening1,broadening2,broadening3} of oscillation frequencies  $|\omega_1({\bf k})-\omega_2({\bf k})|$ of wave vectors ${\bf k}$, and backaction to spin-flips of electrons. This then provides two routes to control the magnon-correlation transport away from the injector: i) the magnetic field causes the spin spatial oscillation in the transport that is enhanced by the applied magnetic field; ii) a proximity conductor brakes the transport, suppresses spatial oscillation, and thereby acts as a gate. Inducing quantum coherence in incoherent processes may lead to crucial applications and new tunability in future magnonic devices.

This article is organized as follows. In Sec.~\ref{general_magnon_spin}, we build up the description of magnon spin by a matrix in general magnetic structures (Sec.~\ref{magnon_spin_general}) and a two-sublattice model of canted AFM (Sec.~\ref{magnon_spin_example}) as a prototypical example. In Sec.~\ref{electric_injection}, we calculate the electrical injection and dephasing of magnon correlation by setting up the quantum kinetic equation. Two functionalities for the spin transport by magnon correlation, including the Hanle effect and gating effect by metals, are addressed in Sec.~\ref{magnon_transport}. 
We summarize our results and give an outlook in Sec.~\ref{summary}.

\section{Spin carried by magnon correlations}
\label{general_magnon_spin}

\subsection{General magnetic structures}

\label{magnon_spin_general}

\begin{figure}[htp!]	
\includegraphics[width=0.41\textwidth,trim=0.0cm 0cm 0cm 0.0cm]{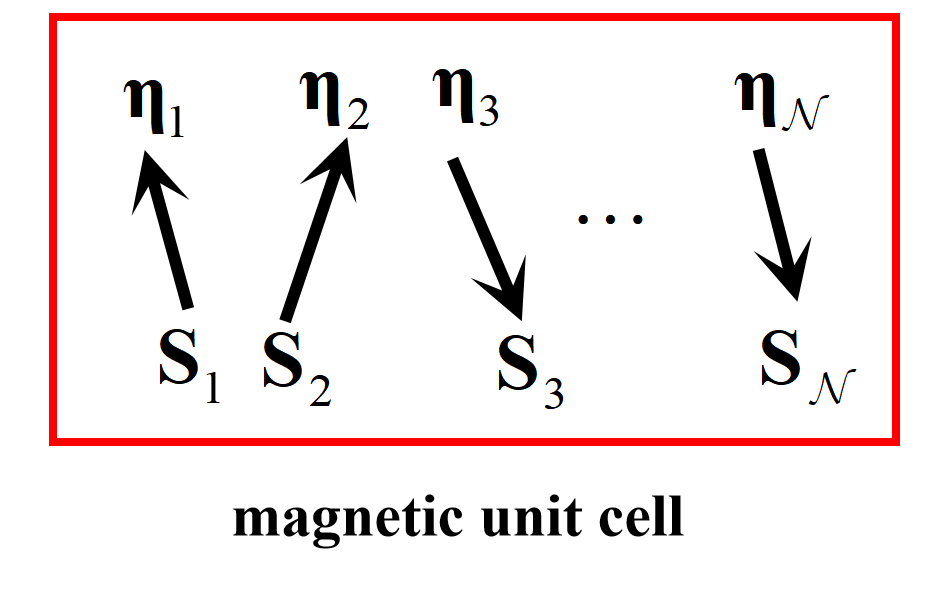}
\caption{A general magnetic structure with ${\cal N}$ noncollinear spins in one unit cell. ${\pmb \eta}_{l=\{1,2\cdots,{\cal N}\}}$ represents their directions in the ground state.}
\label{magnetic_configuration0}
\end{figure}

We start from a general Heisenberg Hamiltonian for noncollinear magnets 
\begin{align}
    \hat{H}_m=(1/2)\sum_{l,l'=1}^{\cal N}\sum_{i,j=1}^{N}\sum_{\alpha,\beta}{J}_{ij}^{\alpha \beta}(l,l')\hat{S}_{li}^{\alpha}\hat{S}_{l'j}^{\beta}
\end{align}
with noncollinear spins $\hat{\bf S}_{li}$ of the $l$-th ion in the $i$-th magnetic unit cell coupled via the coupling constant $J_{ij}^{\alpha \beta}$. The equilibrium configuration of ${\cal N}$ spins with their directions ${\pmb \eta}_{1,\cdots,{\cal N}}$ is illustrated in Fig.~\ref{magnetic_configuration0}.
For the $l$-th magnetic ions in one magnetic unit cell, we define a local $\{\hat{\bf x}'_l,\hat{\bf y}'_l,\hat{\bf z}'_l\}$-frame with $\hat{\bf z}'_l={\pmb \eta}_l$ along its ground-state spin, so the ${\cal N}$ sublattices in one magnetic unit cell with spins along different directions imply ${\cal N}$ local frames.

In the local frames, we represent the spin operator $\hat{\bf S}_{li}$ by the bosonic operators $\hat{a}_{li}$ via the Holstein-Primakoff transformation.
In the linear-response regime, with $|\hat{\bf S}_{li}|=S$, 
\begin{equation}
\begin{aligned}
\hat{S}_{li}^{x'}&=\frac{\sqrt{2 S}}{2}\left(\hat{a}_{li}+\hat{a}_{li}^{\dagger}\right), \\
 \hat{S}_{li}^{y'}&=\frac{\sqrt{2 S}}{2 i}\left(\hat{a}_{li}-\hat{a}_{li}^{\dagger}\right), \\
 \hat{S}_{li}^{z'}&=S-\hat{a}_{li}^{\dagger} \hat{a}_{li}.
\end{aligned}
 \end{equation} 
The local $\{\hat{\bf x}',\hat{\bf y}',\hat{\bf z}'\}$-frame is transformed to the lab $\{\hat{\bf x}, \hat{\bf y},\hat{\bf z} \}$-frame via the rotation matrices $\left(\hat{\bf x}'_l,\hat{\bf y}'_l,\hat{\bf z}'_l\right)$ according to 
\begin{equation}
\left(\begin{array}{c}\hat{S}_{li}^x \\
 \hat{S}_{li}^y \\
\hat{S}_{li}^z\end{array}\right)
=\left({\bf x}'_l,{\bf y}'_l,{\bf z}'_l\right) \left(\begin{array}{c}\hat{S}_{li}^{x'} \\
\hat{S}_{li}^{y'} \\ \hat{S}_{li}^{z'}\end{array}\right),
\label{S-S'}
\end{equation}
such that  the spin operators in the lab coordinate frame read 
\begin{align}
    \hat{\bf S}_{li}=\frac{\sqrt{2S}}{2}{\pmb \xi}_l^{\ast}\hat{a}_{li}+\frac{\sqrt{2S}}{2}{\pmb \xi}_l\hat{a}^{\dagger}_{li}+{\pmb \eta}_l(S-\hat{a}^{\dagger}_{li}\hat{a}_{li}).
    \label{spin}
\end{align}
Here, ${\pmb \xi}_l=\hat{\bf x}'_l+i \hat{\bf y}'_l$ represents the ellipticity of the spin precession and ${\pmb \eta}_l=\hat{\bf z}'_l$ is the direction of ground-state spins.

After the Fourier transformation over the magnetic unit cells $\{i,j\}$, we obtain the Hamiltonian
 \begin{align}
   \hat{H}_m=\frac{1}{2}\sum_{\bf k} \hat{X}_{\bf k}^{\dagger}{\cal H}_{\bf k}\hat{X}_{\bf k},
 \end{align}
in which $\hat{X}_{\bf k}=\left(\hat{a}_{l,{\bf k}}, \\ \hat{a}^{\dagger}_{l,{\bf -k}} \right)^T$ with $\{l=1,...,{\cal N}\}$ and ${\cal H}_{\bf k}$ is the Hamiltonian matrix. In terms of a paraunitary matrix ${\cal T}_{\bf k}$ with ${\cal T}_{\bf k}\sigma_3 {\cal T}_{\bf k}^{\dagger}=\sigma_3$ or ${\cal T}_{\bf k}^{\dagger}\sigma_3 {\cal T}_{\bf k}=\sigma_3$, where 
$\sigma_3=\begin{pmatrix} 
I_{{\cal N} \times {\cal N}} & 0 \\
0 & -I_{{\cal N} \times {\cal N}}
\end{pmatrix}$ 
is the metric matrix, the mode energies  ${\cal E}_{\bf k}\equiv{\rm diag}(\hbar\omega_{1,\bf k},...,\hbar\omega_{{\cal N},\bf k},\hbar\omega_{1,\bf -k},...,\hbar\omega_{{\cal N},\bf -k})={\cal T}_{\bf k}^{\dagger}{\cal H}_{\bf k}{\cal T}_{\bf k}$~\cite{transform_T1}. The eigenvalue problem is solved for the matrix $\sigma_3{\cal H}_{\bf k}$ and the eigenvectors $u_i$ of $\sigma_3 {\cal H}_{\bf k}$ can be arranged to express the transformation matrix ${\cal T}_{\bf k}$.
Accordingly,
\begin{align}
    \hat{H}_m=\frac{1}{2}\sum_{\bf k} \hat{\Psi}_{\bf k}^{\dagger}({\cal T}^{\dagger}_{\bf k}{\cal H}_{\bf k}{\cal T}_{\bf k})\hat{\Psi}_{\bf k}, 
    \label{H_m}
\end{align}
in which $\hat{\Psi}_{\bf k}=\left(\hat{b}_{\xi, {\bf k}}, \\ \hat{b}^{\dagger}_{\xi, {\bf -k}}\right)^T$ defines the magnon operators of modes $\xi=\{1,...,{\cal N}\}$ 
 via
$\hat{X}_{\bf k}={\cal T}_{\bf k}\hat{\Psi}_{\bf k}$.

\begin{figure}[htp!]	
\includegraphics[width=0.55\textwidth,trim=0.0cm 0cm 0cm 0.0cm]{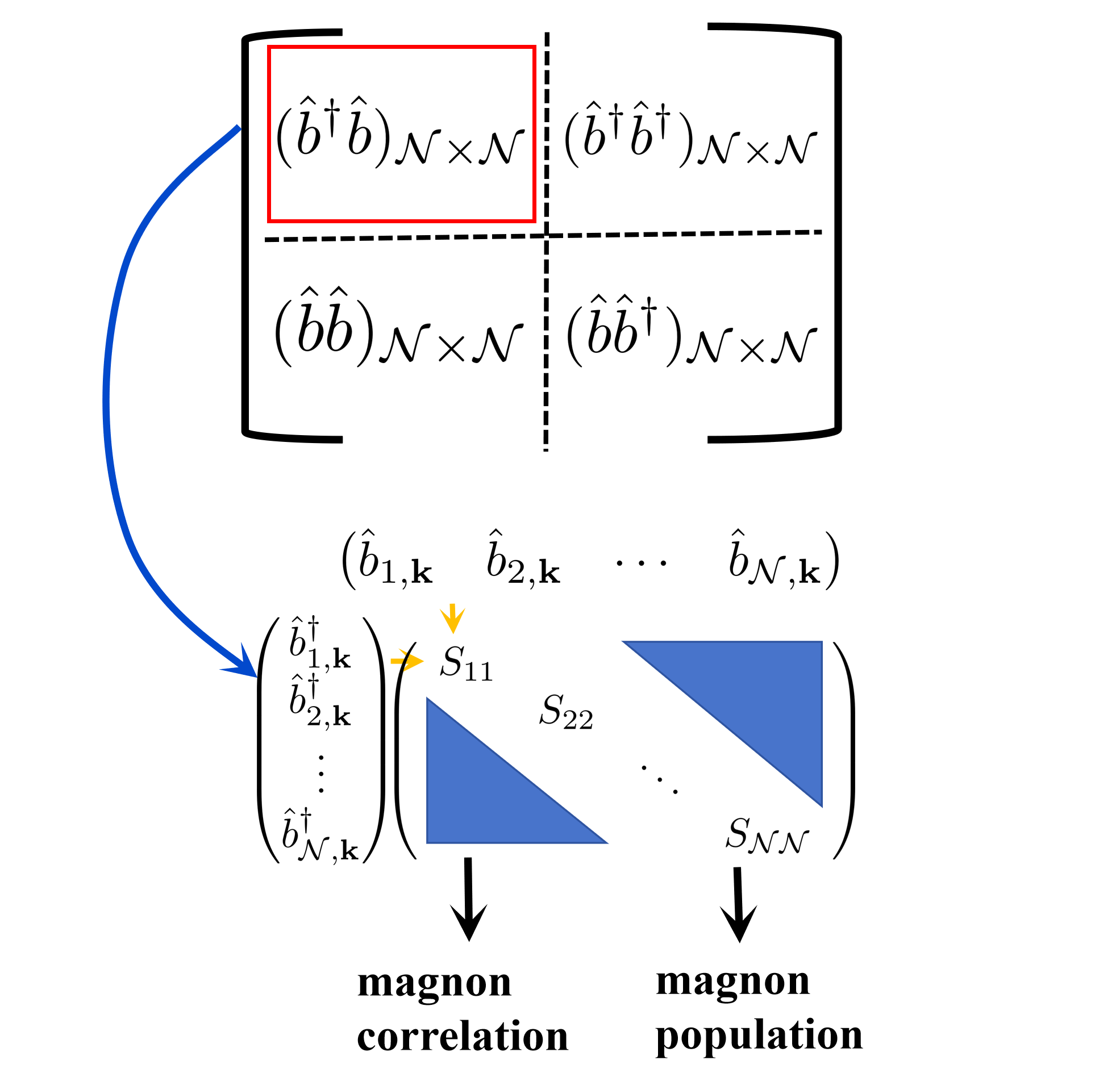}
\caption{The structure of the spin matrix.}
\label{magnetic_configuration}
\end{figure}

\textcolor{blue}{The spin matrix of the magnon is found by rewriting Eq.~\eqref{spin} in the basis of the magnon operators using the Bogoliubov transformation matrix ${\cal T}_{\bf k}$.} 
The $\alpha$-component of the total-spin operator
\begin{align}
\textcolor{blue}{\hat{S}^{\rm tot}_{\alpha}}&=-\frac{1}{2}\sum_{\bf k}\hat{X}^{\dagger}_{\bf k} \Big[I_{2 \times2} \otimes {\rm diag}(\eta^{\alpha}_1,...,\eta^{\alpha}_{\cal N})\Big] \hat{X}_{\bf k}\nonumber\\
&+{\rm zeroth ~ and ~ first ~ order ~ terms ~ of ~}
\{\hat{a},\hat{a}^{\dagger}\},
 \label{total_spin}
 \end{align} 
in which $\eta^{\alpha}_{l = \{1,\cdots,{\cal N}\}}$ is the projection of ${\pmb \eta}_l$ along the $\alpha$-axis of the lab frame. The zeroth and first orders of the bosonic operators $\{\hat{a},\hat{a}^{\dagger}\}$ have no contribution to the magnon spin.
Through the paraunitary matrix ${\cal T}_{\bf k}$, 
\begin{align}
    \hat{S}_{\alpha} =(1/2)\sum_{\bf k}\hat{\Psi}^{\dagger}_{\bf k} {\cal S}_{\alpha}({\bf k}) \hat{\Psi}_{\bf k}
    \label{magnon_spin}
\end{align}
is transformed into the magnon basis with the matrix 
\begin{align}
    {\cal S}_{\alpha}({\bf k})
    &=-{\cal T}^{\dagger}_{\bf k}\left[I_{{2} \times{2}} \otimes {\rm diag}(\eta^{\alpha}_1,...,\eta^{\alpha}_{\cal N})\right]{\cal T}_{\bf k}.
    \label{matrix_spin}
\end{align}
The diagonal elements of ${\cal S}_{\alpha}({\bf k})$ are the magnon spin/polarization defined in the classical description~\cite{local_DMI,spin}. \textcolor{blue}{As illustrated in Fig.~\ref{magnetic_configuration}, the $2{\cal N}\times 2{\cal N}$ spin matrix contains four ${\cal N}\times {\cal N}$ blocks, represented, respectively, by $(\hat{b}^{\dagger}\hat{b})_{{\cal N}\times{\cal N}}$, $(\hat{b}^{\dagger}\hat{b}^{\dagger})_{{\cal N}\times{\cal N}}$, $(\hat{b}\hat{b})_{{\cal N}\times{\cal N}}$, and $(\hat{b}\hat{b}^{\dagger})_{{\cal N}\times{\cal N}}$ blocks. The $(\hat{b}^{\dagger}\hat{b}^{\dagger})_{{\cal N}\times{\cal N}}$ and $(\hat{b}\hat{b})_{{\cal N}\times{\cal N}}$ blocks are magnon spins carried by two creation or annihilation operators. In the $(\hat{b}^{\dagger}\hat{b})_{{\cal N}\times{\cal N}}$ block,} the off-diagonal elements are the magnon correlations; its physical meaning may be fully appreciated in the spin current $\hat{\bf J}_{\alpha}({\bf r})$ and torque $\hat{T}_{\alpha}({\bf r})$ with $\alpha$ denoting the spin polarization.

\textcolor{blue}{Equation~(\ref{matrix_spin}) is defined for any magnetic configuration. There have been specific cases in the literature where the description of the magnon spin matrix is provided. For example, Refs.~\cite{spin2} and \cite{local_DMI} defines the spin $S^{\alpha}_{m,\bf k}=\langle m,{\bf k} |\hat{S}_{\alpha}|m,{\bf k} \rangle -\langle 0 |\hat{S}_{\alpha}|0\rangle$ carried by the magnon mode $m=\{1,...,{\cal N}\}$, which only addresses the diagonal terms $[{\cal S}_{\alpha}({\bf k})]_{mm}$ of the spin matrix, as shown in Fig.~\ref{magnetic_configuration}. These spins are carried by the magnon population that is responsible for the spin Seebeck effect. On the other hand, Ref.~\cite{spin3} investigated the excitation of Brillouin zone-edge magnon-pairs ($\hat{b}^{\dagger}_{l}\hat{b}^{\dagger}_{l'}$ with $l\neq l'$) using optical pulses. In their subsequent work, Ref.~\cite{spin4} discussed ultrafast, optically-controlled spin currents carried by such magnon pairs in collinear antiferromagnets.}

\textcolor{blue}{The matrix description represents the real spin in the sense that it originates from the magnetic moments of every ion, as in Fig.~\ref{magnetic_configuration0}. 
For a magnetic unit cell with ${\cal N}$ sublattices, the creation of a Holstein-Primakoff boson decreases the spin $\hbar$  along the magnetic moment of that ion. The magnon operator connects to the Holstein-Primakoff boson via the Bogoliubov transformation, which then carries the real spin of the system. This microscopic perspective distinguishes itself from other discussions of angular momentum. For instance, Ref.~\cite{spin5} focuses on ferromagnets, which include the orbital angular momentum that is absent in our canted antiferromagnets. Reference~\cite{spin6} addresses the pseudo-orbital angular momentum of phonons and magnons, while we focus on spin angular momentum. Reference~\cite{easy_plane8} constructs the pseudospin $(\tilde{L}_x,\tilde{L}_y,\tilde{L}_z)$, which is only a convenience to expand a $2\times 2$ Hamiltonian matrix of an antiferromagnet by the  Holstein-Primakoff bosons. In such a peusospin description, only $\tilde{L}_z$ represents the real spin of the magnon along the N\'eel vector, while $\tilde{L}_x$ and $\tilde{L}_y$ do not correspond to measurable spins.}

\textcolor{blue}{The off-diagonal magnon correlations contribute to the spin matrix when the associated spin component is not conserved, i.e., when $[\hat{S}_{\alpha},\hat{H}_m]\neq 0$ due to broken spin rotational symmetry. By substituting Eqs.~(\ref{H_m}) and (\ref{magnon_spin}), we find
\begin{align}
   [\hat{S}_{\alpha},\hat{H}_m]=\frac{1}{2}\sum_{{\bf k}}
    \hat{\Psi}^{\dagger}_{\bf k}[{\cal S}_{\alpha}({\bf k})\sigma_3 {\cal E}_{\bf k}-{\cal E}_{\bf k}\sigma_3 {\cal S}_{\alpha}({\bf k})]\hat{\Psi}_{\bf k}.
    \label{commutation_relation}
\end{align}
Accordingly, when  $[\hat{S}_{\alpha},\hat{H}_m]=0$ and ${\cal E}_{\bf k}$ is non-degenerate, the spin matrix ${\cal S}_{\alpha}({\bf k})$ should be diagonal, i.e., the magnon correlation contributes no spin operator. In other words, when $[\hat{S}_{\alpha},\hat{H}_m]\neq 0$, the off-diagonal components in the spin matrix exist and contribute to the magnon spin. In collinear systems, such as conventional ferromagnets and easy-axis antiferromagnets, where $U(1)$ spin symmetry around the easy-axis is preserved, the spin matrix is diagonal, such that the diagonal terms solely describe spin. Conversely, in noncollinear and canted antiferromagnets, the $U(1)$ symmetry can be broken, rendering the off-diagonal correlation terms a contribution to the spin. }

The spin current and spin torque densities of magnons are derived from the continuity equation of the spin-density operator 
\begin{align}
    \hat{S}_{\alpha}({\bf r})&=-(1/2)\hat{X}^{\dagger}({\bf r}) \left[I_{{2} \times{2}} \otimes {\rm diag}(\eta^{\alpha}_1,...,\eta^{\alpha}_{\cal N})\right] \hat{X}({\bf r})\nonumber\\
    &=\frac{1}{V}\sum_{\bf q}\hat{D}_{\alpha}({\bf q})e^{i{\bf q}\cdot{\bf r}}.
\end{align}
The Fourier component of the spin-density operator reads
\begin{align}
    \hat{D}_{\alpha}({\bf q})&=-\frac{1}{2}\sum_{\bf k}\hat{X}^{\dagger}_{\bf k}\left[I_{{2} \times{2}} \otimes {\rm diag}(\eta^{\alpha}_1,...,\eta^{\alpha}_{\cal N})\right]\hat{X}_{\bf k+q} \nonumber\\
    &=\frac{1}{2}\sum_{\bf k}\hat{\Psi}^{\dagger}_{\bf k}{\cal D}_{\alpha}({\bf k},{\bf q})\hat{\Psi}_{\bf k+q},
    \label{S_q}
\end{align}
where the matrix 
\[
{\cal D}_{\alpha}({\bf k},{\bf q})\equiv -{\cal T}^{\dagger}_{\bf k}\left[I_{{2} \times{2}} \otimes {\rm diag}(\eta^{\alpha}_1,...,\eta^{\alpha}_{\cal N})\right]{\cal T}_{\bf k+q}.
\]
In the continuity equation
\begin{align}
    \frac{\partial \hat{S}_{\alpha}({\bf r})}{\partial t}=-\nabla \cdot \hat{\bf J}_{\alpha}({\bf r})+\hat{T}_{\alpha}({\bf r}),
    \label{continuity_1}
\end{align}
$\hat{\bf J}_{\alpha}({\bf r})$ and $\hat{T}_{\alpha}({\bf r})$ represent the spin-current and spin-torque densities.
According to $\nabla \cdot \hat{\bf J}_{\alpha}({\bf r})=(1/V)\sum_{\bf q} i{\bf q}\cdot \hat{\bf J}_{\alpha}({\bf q})e^{i{\bf q}\cdot {\bf r}}$, Eq.~\eqref{continuity_1} implies 
\begin{align}
    \frac{\partial \hat{D}_{\alpha}({\bf q})}{\partial t}=-i{\bf q}\cdot \hat{\bf J}_{\alpha}({\bf q})+\hat{T}_{\alpha}({\bf q}).
    \label{continue_Eq}
\end{align}
On the other hand, by the Heisenberg equation of motion,
\[
\frac{\partial \hat{D}_{\alpha}({\bf q})}{\partial t}=\frac{1}{i\hbar}\left(\left[\hat{D}_{\alpha}({\bf q}),\hat{H}_m\right]+\left[\hat{D}_{\alpha}({\bf q}),\hat{H}_{\rm int}\right]\right),
\]
where $\hat{H}_{\rm int}$ originates from the interplay with environment. 
The second term contributes an external spin torque $\hat{T}_{\rm ext}$ induced by interactions with the environment. We are concerned with the intrinsic contribution, which leads to the spin-current operator and the \textit{intrinsic} spin torque $\hat{T}_{\rm in}$.

Based on the commutators between the magnon operators $[\hat{b}_{\xi,{\bf k}},\hat{b}^{\dagger}_{\xi',\bf k'}]=\delta_{\xi \xi'}\delta_{\bf kk'}$ and $[\hat{b}_{\xi,{\bf k}},\hat{b}_{\xi',\bf k'}]=0$, the field operators $\hat{\Psi}_{\bf k}=\left(\hat{b}_{1, {\bf k}}, \\ \hat{b}_{2, {\bf k}},\\\hat{b}^{\dagger}_{1, {\bf -k}},\\ \hat{b}^{\dagger}_{2, {\bf -k}}\right)^T$ obey $[\hat{\Psi}_{n,{\bf k}},\hat{\Psi}^{\dagger}_{m,{\bf k'}}]=(\sigma_3)_{nm}\delta_{\bf kk'}$, $[\hat{\Psi}_{n,{\bf k}},\hat{\Psi}_{m,{\bf k'}}]=(i\sigma_2)_{nm}\delta_{{\bf -k},{\bf k'}}$, and $[\hat{\Psi}^{\dagger}_{n,{\bf k}},\hat{\Psi}^{\dagger}_{m,{\bf k'}}]=(-i\sigma_2)_{nm}\delta_{{\bf -k},{\bf k'}}$, 
where $\sigma_2=\begin{pmatrix} 
0 & -iI_{{\cal N} \times {\cal N}} \\
iI_{{\cal N} \times {\cal N}} & 0
\end{pmatrix}$. 
By combining Eqs.~(\ref{H_m}) and (\ref{S_q}), and with the diagonal matrix ${\cal E}_{\bf k}={\cal T}_{\bf k}^{\dagger}{\cal H}_{\bf k}{\cal T}_{\bf k}$, 
we find
\begin{align}
   &[\hat{D}_{\alpha}({\bf q}),\hat{H}_m]\nonumber\\
    &=\frac{1}{2}\sum_{{\bf k}}
    \hat{\Psi}^{\dagger}_{\bf k}\left({\cal D}_{\alpha}({\bf k},{\bf q})\sigma_3 {\cal E}_{\bf k+q}-{\cal E}_{\bf k}\sigma_3 {\cal D}_{\alpha}({\bf k},{\bf q})\right)\hat{\Psi}_{\bf k+q}.\nonumber\\
\end{align}
When the external driven field varies slowly in the real space, we participate only the excitation of small wave vector ${\bf q}$ (long wavelength) is involved. 
By shifting ${\bf k} \rightarrow {\bf k}-{\bf q}/2$, to the leading order of ${\bf q}$, we obtain the spin-current density operator~\cite{current2},
\begin{align}
    \hat{\bf J}_{\alpha}({\bf q})&=\frac{1}{4}\sum_{{\bf k}}
    \hat{\Psi}^{\dagger}_{{\bf k}-{\bf q}/2}({\cal D}_{\alpha}({\bf k}-{\bf q}/2,{\bf q})\sigma_3{\pmb {\cal V}}_{\bf k}\nonumber\\
    &+{\pmb {\cal V}}_{\bf k}\sigma_3 {\cal D}_{\alpha}({\bf k}-{\bf q}/2,{\bf q}))\hat{\Psi}_{{\bf k}+{\bf q}/2},
    \label{J_o}
\end{align}
and the intrinsic spin-torque density operator 
\begin{align}
    \hat{T}^{\alpha}_{{\rm in}}({\bf q})&=\frac{1}{2i\hbar}\sum_{{\bf k}}
    \hat{\Psi}^{\dagger}_{{\bf k}-{\bf q}/2}({\cal D}_{\alpha}({\bf k}-{\bf q}/2,{\bf q})\sigma_3 {\cal E}_{\bf k}\nonumber\\
    &- {\cal E}_{\bf k}\sigma_3 {\cal D}_{\alpha}({\bf k}-{\bf q}/2,{\bf q}))\hat{\Psi}_{{\bf k}+{\bf q}/2},
\end{align}
in which magnon group velocities  ${\pmb {\cal V}}_{{\bf k}}=(1/\hbar)\partial {\cal E}_{\bf k}/\partial {\bf k}$. 
The non-vanishing spin torque density indicates the non-conservation of the injected spin.

When flowing uniformly, ${\bf q}=0$ in Eq.~(\ref{J_o}), such that the spin current and torque read~\cite{current1,current2}
\begin{subequations}
\begin{align}
   \hspace{-0.3cm} \hat{\bf J}_{0\alpha}&=\frac{1}{4V}\sum_{{\bf k}}
    \hat{\Psi}^{\dagger}_{{\bf k}}\left({\cal S}_{\alpha}({\bf k})\sigma_3{\pmb  {\cal V}}_{\bf k}+{\pmb {\cal V}}_{\bf k}\sigma_3 {\cal S}_{\alpha}({\bf k})\right)\hat{\Psi}_{{\bf k}},
    \label{J_s0}\\
     \hspace{-0.3cm}\hat{T}_{0\alpha}&=\frac{1}{2V}\frac{1}{i\hbar}\sum_{{\bf k}}
    \hat{\Psi}^{\dagger}_{{\bf k}}\left({\cal S}_{\alpha}({\bf k})\sigma_3 {\cal E}_{\bf k}- {\cal E}_{\bf k}\sigma_3 {\cal S}_{\alpha}({\bf k})\right)\hat{\Psi}_{{\bf k}}.
    \label{torque}
\end{align}
\end{subequations}

\subsection{Two-sublattice model of AFM}

\label{magnon_spin_example}

The magnon polarization/spin in the classical description is often quenched by specific interactions since they hybridize two circularly polarized modes, such as local Dzyaloshinskii-Moriya interaction~\cite{local_DMI,local_DMI2,local_DMI3} or transverse magnetic field ${\bf H}$, rendering AFMs canted. Using a two-sublattice AFM as a prototypical model, we illustrate the spin matrix and its physical meaning.

\begin{figure}[htp!]	
\includegraphics[width=0.48\textwidth,trim=0.6cm 0cm 0cm 0.1cm]{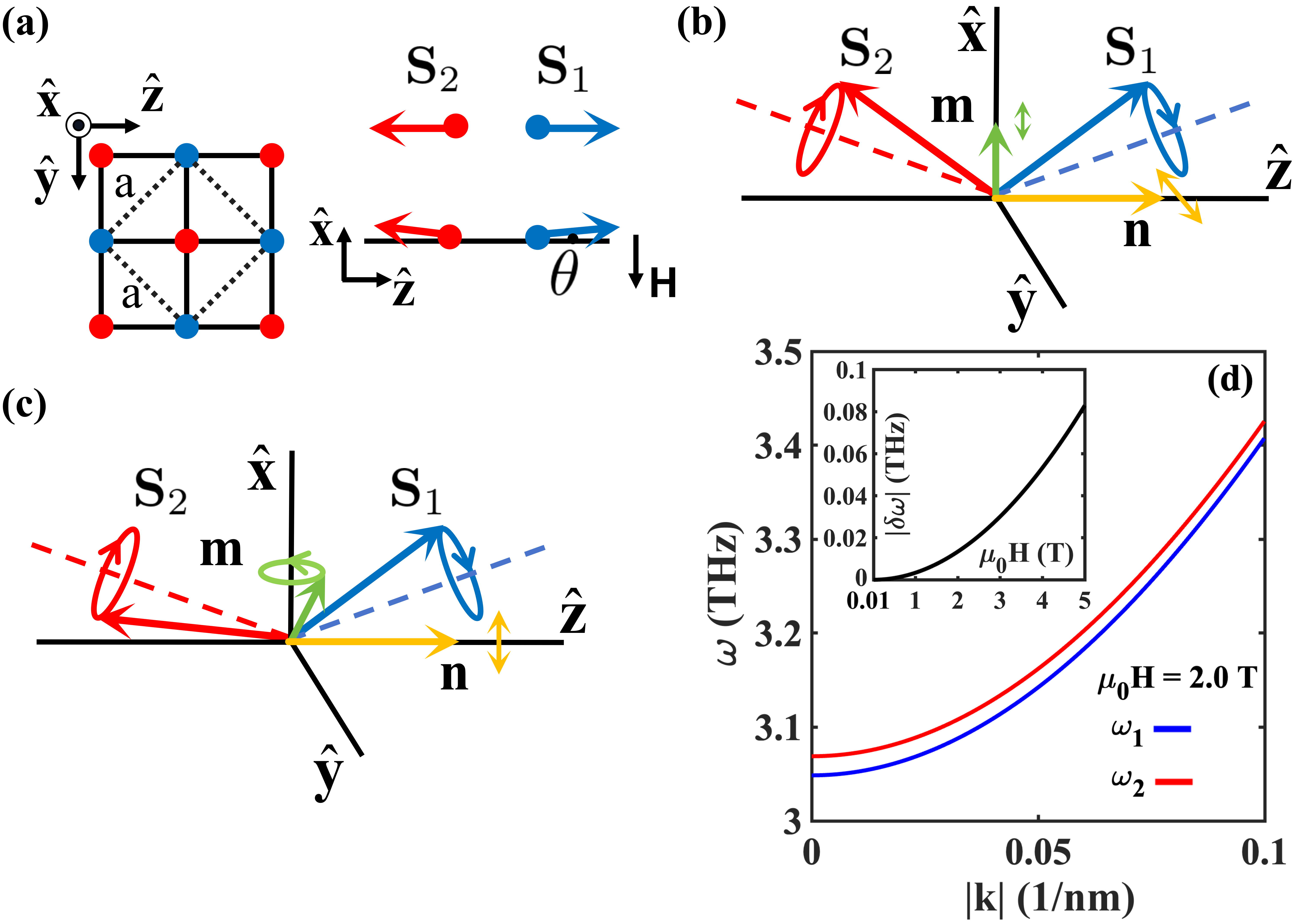}
\caption{(a) Two-sublattice model of AFM. The dashed square represents a unit cell with a lattice constant $a$. (b) and (c) Polarizations ${\bf m}$ by green arrows of the anti-phase mode ``$1$" and in-phase mode ``$2$" with the N\'eel vector ${\bf n}$ of mode ``1" (``2") oscillating along the $\hat{\bf y}~(\hat{\bf x})$-axis, indicated by orange arrows. (d) Band structure of two modes. Inset: average frequency difference $|\delta\omega|=|\omega_1-\omega_2|$ over $|{\bf k}|\in[0,0.5]$~nm$^{-1}$.}
\label{model}
\end{figure}

As depicted in Fig.~\ref{model}(a), two nearest-neighboring spins couple via the antiferromagnetic exchange interaction $J>0$ and are aligned along the N\'eel vector $\hat{\bf z}$-axis by the easy-axis anisotropy $K_z<0$, governed by the Hamiltonian 
\begin{align}
\hat{H}_m&=J\sum_{\langle i,j\rangle} \hat{\bf S}_{1i} \cdot \hat{\bf S}_{2j}+K_z\sum_l^{2}\sum_{i}(\hat{S}_{li}^z)^2\nonumber\\
&-|\gamma|\hbar\mu_0 H \sum_l^{2}\sum_{i} \hat{S}^x_{li},
\label{1D_H0}
\end{align} 
where $\langle \cdots \rangle$ represents the summation over the nearest-neighboring sites $\{i,j\}$ and $\gamma<0$ is the gyromagnetic ratio of electrons. 
A magnetic field ${\bf H}$ along the $-\hat{\bf x}$-direction stabilizes a canted spin configuration with $\hat{\bf S}_1=S(\sin \theta, 0, \cos \theta)$ and $\hat{\bf S}_2=S(\sin \theta, 0, -\cos \theta)$ not parallel in one magnetic unit cell by a canted angle $\theta={\rm arcsin}\left(|\gamma|\hbar\mu_0H/(8JS-2K_zS)\right)$.
In the lab coordinate, ${\bf S}_{1,2}$ are along ${\pmb \eta}_{1(2)}=(\sin\theta,0, \pm \cos\theta)^T$.
There are two modes with the field operator of magnons $\hat{\Psi}_{{\bf k}}=(\hat{b}_{1,{\bf k}},\hat{b}_{2,{\bf k}},\hat{b}^{\dagger}_{1,-{\bf k}},\hat{b}^{\dagger}_{2,-{\bf k}})^T$. 

According to Eq.~(\ref{spin}), 
the Hamiltonian $\hat{H}_m = (1/2)\sum_{\bf k} \hat{X}^{\dagger}_{\bf k}  {\cal H}_{\bf k} \hat{X}_{\bf k}$ in terms of the bosonic operators $\hat{X}_{\bf k}=(\hat{a}_{1,{\bf k}},\hat{a}_{2,{\bf k}},\hat{a}^{\dagger}_{1,-{\bf k}},\hat{a}^{\dagger}_{2,-{\bf k}})^T$, where the Hamiltonian matrix 
\begin{align}
    {\cal H}({\bf k})=S\begin{pmatrix}
        A & C_{+}({\bf k}) & B & C_{-}({\bf k}) \\
        C_{+}({\bf k}) & A & C_{-}({\bf k}) & B \\
        B & C_{-}({\bf k}) & A & C_{+}({\bf k}) \\
        C_{-}({\bf k}) & B & C_{+}({\bf k}) & A
    \end{pmatrix}.
    \label{H_x}
\end{align}
Here $S=|{\bf S}|$, $A=4J\cos2\theta+|\gamma|\hbar\mu_0 H\sin{\theta}/S+K_z(\sin^2\theta-2\cos^2\theta)$, $B=(1/2)K_z\sin^2\theta$, and $C_{\pm}({\bf k})=\pm J(1 \mp \cos 2\theta)(\cos(\sqrt{2}k_ya/2)+\cos(\sqrt{2}k_za/2))$ with $a$ denoting the lattice constant.
The dispersion of the two modes 
\begin{align}
    \hbar \omega_{1}({\bf k})&=S\sqrt{(A-C_{+}({\bf k}))^2-(B-C_{-}({\bf k}))^2}, \nonumber \\
    \hbar \omega_{2}({\bf k})&=S\sqrt{(A+C_{+}({\bf k}))^2-(B+C_{-}({\bf k}))^2},
    \label{E_k}
\end{align}
are two positive eigenvalues of $\sigma_3 {\cal H}({\bf k})$, where the metric $\sigma_3={\rm diag}\{I_{2\times 2},-I_{2\times 2}\}$. 
 We adopt a hyperbolic parametrization in terms of the parameters $\{\varphi,\phi,m,l\}$: $A-C_{+}=m\cosh{\varphi}$, $A+C_{+}=l\cosh{\phi}$, $B-C_{-}=m\sinh{\varphi}$, and $B+C_{-}=l\sinh{\phi}$ such that $\hbar \omega_{1}=Sm$ and $\hbar \omega_{2}=Sl$. The corresponding eigenvectors of $\sigma_3 {\cal H}({\bf k})$ form the Bogoliubov transformation matrix 
\begin{align}
    {\cal T}({\bf k})=\frac{\sqrt{2}}{2}\begin{pmatrix}
        -{\cal T}_{1,{\bf k}}  &  {\cal T}_{2,{\bf k}}  &  {\cal T}_{1,{\bf k}}' & -{\cal T}_{2,{\bf k}}' \\
         {\cal T}_{1,{\bf k}}  &  {\cal T}_{2,{\bf k}}  & -{\cal T}_{1,{\bf k}}' & -{\cal T}_{2,{\bf k}}'\\
         {\cal T}_{1,{\bf k}}' & -{\cal T}_{2,{\bf k}}' & -{\cal T}_{1,{\bf k}}  &  {\cal T}_{2,{\bf k}}\\
        -{\cal T}_{1,{\bf k}}' & -{\cal T}_{2,{\bf k}}' &  {\cal T}_{1,{\bf k}}  &  {\cal T}_{2,{\bf k}}
    \end{pmatrix},
    \label{Tmatrix_0}
\end{align} 
where 
\begin{align}
  &{\cal T}_{1,{\bf k}}=\cosh({\varphi}/{2})=(1/\sqrt{2})\sqrt{[A-C_{+}({\bf k})]S/\hbar\omega_{1,{\bf k}}+1},\nonumber\\
  &{\cal T}_{1,{\bf k}}'=\sinh({\varphi}/{2})=(1/\sqrt{2})\sqrt{[A-C_{+}({\bf k})]S/\hbar\omega_{1,{\bf k}}-1},\nonumber\\
  &{\cal T}_{2,{\bf k}}=\cosh({\phi}/{2})=(1/\sqrt{2})\sqrt{[A+C_{+}({\bf k})]S/\hbar\omega_{2,{\bf k}}+1},\nonumber\\
  &{\cal T}_{2,{\bf k}}'=\sinh({\phi}/{2})=-(1/\sqrt{2})\sqrt{[A+C_{+}({\bf k})]S/\hbar\omega_{2,{\bf k}}-1}. \nonumber 
\end{align}

Figure~\ref{model}(b) and (c) plot the spin precession of the two modes: the precession of ${\bf S}_1$ and ${\bf S}_2$ is out of phase in mode ``$1$", whereas in mode ``$2$" it is in phase. Both modes have vanished polarization along the N\'eel vector ${\bf n}$.
Figure~\ref{model}(d) addresses the band structures $\omega_{1,2}({\bf k})$ at $\mu_0H=2$~T, calculated with $J=4$~meV and $K_z=-0.01$~meV; the inset shows enhanced average frequency difference $|\delta\omega|=|\omega_1-\omega_2|$ by magnetic fields.

According to Eqs.~(\ref{matrix_spin}) and (\ref{Tmatrix_0}), the total spin operator $\hat{S}_{z} =({1}/{2})\sum_{{\bf k}}\hat{\Psi}^{\dagger}_{{\bf k}} {\cal S}_{z}({\bf k}) \hat{\Psi}_{{\bf k}}$ along the N{\'e}el vector with  the field operator of magnons $\hat{\Psi}_{{\bf k}}=(\hat{b}_{1,{\bf k}},\hat{b}_{2,{\bf k}},\hat{b}^{\dagger}_{1,-{\bf k}},\hat{b}^{\dagger}_{2,-{\bf k}})^T$, in which the spin matrix 
\begin{align}
    {\cal S}_z({\bf k})=\begin{pmatrix}
        0 & {\cal A}_{{\bf k}} & 0 & {\cal B}_{{\bf k}} \\
        {\cal A}_{{\bf k}} & 0 & {\cal B}_{{\bf k}} & 0 \\
        0 & {\cal B}_{{\bf k}} & 0 & {\cal A}_{{\bf k}}\\
        {\cal B}_{{\bf k}} & 0 & {\cal A}_{{\bf k}} & 0
    \end{pmatrix}
\end{align}
with 
${\cal A}_{{\bf k}}=({\cal T}_{1,{\bf k}}{\cal T}_{2,{\bf k}}+{\cal T}_{1,{\bf k}}'{\cal T}_{2,{\bf k}}')\cos\theta$ and ${\cal B}_{{\bf k}}=-({\cal T}_{1,{\bf k}}{\cal T}_{2,{\bf k}}'+{\cal T}_{2,{\bf k}}{\cal T}_{1,{\bf k}}')\cos\theta$. Accordingly, the total spin operator along the N{\'e}el vector 
\begin{align}
    \hat{S}_{z} =\sum_{{\bf k}}\left({\cal A}_{\bf k}\hat{b}_{1,{\bf k}}^{\dagger}\hat{b}_{2,{\bf k}}+{\cal B}_{\bf k}\hat{b}_{1,{\bf k}}^{\dagger}\hat{b}^{\dagger}_{2,{-\bf k}}\right)+{\rm H.c.}.
    \label{Sz}
\end{align}
The population of both magnon modes carries no spin along the N\'eel vector; the spin is solely carried by the correlations between the two modes.

The velocity of magnon ${\pmb {\cal V}}_{{\bf k}}={\rm diag}({\bf v}_{1,{\bf k}},{\bf v}_{2,{\bf k}},-{\bf v}_{1,-{\bf k}},-{\bf v}_{2,-{\bf k}})$, where ${\bf v}_{n,{\bf k}}=\partial \omega_{n,{\bf k}}/\partial {\bf k}$. According to Eq.~(\ref{J_s0}), we obtain the uniform spin-current density operator with the spin polarization along the N{\'e}el vector $\hat{\bf z}$-direction:
\begin{align}
    \hat{J}_{0z}&=\frac{1}{2V}\sum_{{\bf k}}
    ({\cal A}_{\bf k}({\bf v}_{1,{\bf k}}+{\bf v}_{2,{\bf k}})\hat{b}_{1,{\bf k}}^{\dagger}\hat{b}_{2,{\bf k}}\nonumber\\
    &+{\cal B}_{\bf k}({\bf v}_{1,{\bf k}}+{\bf v}_{2,-{\bf k}})\hat{b}_{1,{\bf k}}^{\dagger}\hat{b}^{\dagger}_{2,{-\bf k}})+{\rm H.c.}.
    \label{spin_current_2D}
\end{align}
Since ${\cal H}_{\bf k}={\cal H}_{-\bf k}$, ${\bf v}_{2,-\bf k}=-{\bf v}_{2,\bf k}$, such that with ${\bf v}_{1,\bf k} \approx {\bf v}_{2,\bf k}$ only the first term in Eq.~\eqref{spin_current_2D} contributes, implying that the magnon correlation carries the magnon spin current but not the magnon population. 
Furthermore, due to the breaking of the $U(1)$ symmetry in canted AFMs, $S_z$ is no longer a conserved quantity. The intrinsic torque by Eq.~(\ref{torque})  
\begin{align}
    \hat{T}_{0z}&=\frac{1}{V}\sum_{{\bf k}}
    (i{\cal A}_{\bf k}(\omega_{1,{\bf k}}-\omega_{2,{\bf k}})\hat{b}_{1,{\bf k}}^{\dagger}\hat{b}_{2,{\bf k}}\nonumber\\
    &+i{\cal B}_{\bf k}(\omega_{1,{\bf k}}+\omega_{2,-{\bf k}})\hat{b}_{1,{\bf k}}^{\dagger}\hat{b}^{\dagger}_{2,{-\bf k}})+{\rm H.c.}
\end{align}
is also carried by the magnon correlation.

A solution ${\cal A}_{\bf k}\approx 1$ and ${\cal B}_{\bf k}\approx 0$ when the canted angle is small with $|\gamma|\hbar\mu_0H \ll 8JS$ is sufficient to analyze the mechanism. The spin current and torque for spin polarization along ${\bf n}$ are also solely carried by magnon correlations.
Indeed, according to Eqs.~(\ref{J_s0}) and (\ref{torque}), 
\begin{subequations}
\begin{align}
    \hat{J}_{0z}&\approx\frac{1}{2V}\sum_{{\bf k}}
    {\cal A}_{\bf k}({\bf v}_{1,{\bf k}}+{\bf v}_{2,{\bf k}})\hat{b}_{1,{\bf k}}^{\dagger}\hat{b}_{2,{\bf k}}+{\rm H.c.},
    \label{Jsz}\\
    \hat{T}_{0z}&\approx \frac{1}{V}\sum_{{\bf k}}
    i{\cal A}_{\bf k}(\omega_{1,{\bf k}}-\omega_{2,{\bf k}})\hat{b}_{1,{\bf k}}^{\dagger}\hat{b}_{2,{\bf k}}+{\rm H.c.},
\end{align}
\end{subequations}
without contribution from the magnon population.

\section{Electrical injection of magnon correlation}

\label{electric_injection}

\subsection{General magnetic structures}

We examine the electrical spin injection from heavy metals to noncollinear magnets in a bilayer structure as illustrated in Fig.~\ref{inj_model}(b). 
When a charge current is applied along the $\hat{\bf y}$-direction in the metals, the injected spin current along $\hat{\bf x}$ generates a spin accumulation $\mu_{\uparrow}\neq \mu_{\downarrow}$ polarized along the N\'eel vector $\hat{\bf z}$ at the interface through the spin Hall effect.
This accumulated spin of electrons then excites magnon correlations in the underlying AFM as in Fig.~\ref{inj_model}(b) via the $s$-$d$ exchange interaction between local spins and conduction electrons at the interface, governed by the Hamiltonian~\cite{s_d_Ham,interface20,interface7,Ping_Tang}
\begin{align}
    \hat{H}_{\rm int}=\sum_{i \in {\rm int}}^{N_{\rm int}} \sum_l^{\cal N} {\cal J}_l \hat{\bf s}_i \cdot \hat{\bf S}_{li},
\end{align}
where the summation is performed over the magnetic unit cells ``$i$",
with $N_{\rm int}$ being their total number in the interfacial plane, and the spin sublattices ``$l$".
${\cal J}_l$ represents the local exchange coupling, which typically depends on the AFM sublattices. Equal or unequal coupling with different sublattices corresponds to compensated and uncompensated interfaces, as observed experimentally~\cite{interface21,interface22,interface23}. 

The electrons in conductors are described in the lattice model with $M$ lattices and the lattice constant $a$. The electron spin at the lattice 
\begin{align}
    \hat{\bf s}_i=\sum_{\alpha,\beta} {\pmb \sigma}_{\alpha \beta}\hat{c}^{\dagger}_{\alpha,i}\hat{c}_{\beta,i}=\frac{1}{M}\sum_{\alpha,\beta}\sum_{{\bf k},{\bf k'}} {\pmb \sigma}_{\alpha \beta}\hat{c}^{\dagger}_{\alpha,{\bf k}-{\bf k'}}\hat{c}_{\beta,{\bf k}}e^{i{\bf k'}\cdot {\bf r}_i},
    \label{lattice_spin}
\end{align}
where ${\pmb \sigma}$ is the Pauli matrices and $\hat{c}_{\beta,i}=(1/\sqrt{M})\sum_{{\bf k}}\hat{c}_{{\bf k},\beta}e^{i{{\bf k}}\cdot {\bf r}_i}$ is the annihilation operator of electrons. 
On the other hand, the electron spin density $\hat{\bf s}({\bf r})=\sum_{\alpha,\beta} {\pmb \sigma}_{\alpha \beta}\hat{c}^{\dagger}_{\alpha}({\bf r})\hat{c}_{\beta}({\bf r})$. 
With $\hat{c}_{\beta}({\bf r})=(1/\sqrt{V})\sum_{{\bf k}}\hat{c}_{{{\bf k}},\beta}e^{i{{\bf k}}\cdot {\bf r}}$, where $V=Ma^3=N_{\rm int}a^2d$ is the volume of the conductors of thickness $d$, we obtain
\begin{align}
    \hat{\bf s}({\bf r})=\frac{1}{Ma^3}\sum_{\alpha,\beta}\sum_{{\bf k},{\bf k'}} {\pmb \sigma}_{\alpha \beta}\hat{c}^{\dagger}_{\alpha,{{\bf k}-{\bf k'}}}\hat{c}_{\beta,{{\bf k}}}e^{i{{\bf k'}}\cdot {\bf r}}.
    \label{spin_density}
\end{align}
By comparing Eqs.~\eqref{lattice_spin} and \eqref{spin_density}, we find $\hat{\bf s}_i=a^3\int d{\bf r} \delta({\bf r}-{\bf r}_i) \hat{\bf s}({\bf r})=a^3\hat{\bf s}({\bf r}_i)$.
In terms of the spin density operator $\hat{\bf s}({\bf r})$, 
\begin{align}
    \hat{H}_{\rm int}=\sum_{i \in {\rm int}}^{N_{\rm int}} \sum_l^{\cal N}{\cal J}_l a^3\int d{\bf r} \delta({\bf r}-{\bf r}_i) \hat{\bf s}({\bf r}) \cdot \hat{\bf S}_{li}.
    \label{Ham}
\end{align}

With the local spin operator $\hat{\bf S}_{li}$ in Eq.~\eqref{spin} and $\hat{a}_{li}=(1/\sqrt{N_{\rm int}})\sum_{\bf q}\hat{a}_{l,{\bf q}}e^{i{\bf q}\cdot {\bf r}_i}$, the Hamiltonian \eqref{Ham} becomes 
\begin{align}
    \hat{H}_{\rm int}
    &=\frac{a}{d}\sqrt{\frac{S}{2N_{\rm int}}}\sum_{{\bf k},{\bf k}',{\bf q}}\sum_{l=1}^{\cal N}\sum_{\alpha \beta} {\cal J}_l {\pmb \sigma}_{\alpha \beta} \cdot {\pmb \xi}_l^{\ast}
    \hat{c}^{\dagger}_{{{\bf k}'},\alpha}\hat{c}_{{{\bf k}},\beta}\hat{a}_{l,{\bf q}}\delta_{{{\bf k}}+{\bf q}, {{\bf k}'}}\nonumber\\
    &+{\rm H.c.}.
    \label{Ham_process}
\end{align}
The bosonic operator $\hat{a}$ is expressed in terms of the magnon operator $\hat{b}$ by the Bogoliubov transformation 
\begin{align}
    \hat{a}_{l,{\bf q}}=\sum_{\zeta=1}^{\cal N}\left({\cal T}_{l,\zeta}({\bf q})\hat{b}_{\zeta,{\bf q}} +{\cal T}_{l,\zeta+{\cal N}}({\bf q})\hat{b}^{\dagger}_{\zeta,{\bf -q}}\right), 
    \label{ab}
\end{align}
where ${\cal T}({\bf q})$ is the Bogoliubov transformation matrix.
Substituting Eq.~\eqref{ab} into (\ref{Ham_process}) leads to
\begin{align}
    \hat{H}_{\rm int}&=\frac{a}{d}\sqrt{\frac{S}{2N_{\rm int}}}\sum_{{\bf k},{\bf k}',{\bf q}}\sum_{\zeta=1}^{\cal N}\sum_{\alpha \beta}
    {\pmb \sigma}_{\alpha \beta} \cdot {\pmb \chi}^{(\zeta)}_{\bf q}\hat{c}_{{{\bf k}'},\alpha}^{\dagger}\hat{c}_{{{\bf k}},\beta}\hat{b}_{\zeta,{\bf q}}\delta_{{{\bf k}}+{\bf q},{\bf k}'}\nonumber\\
    &+{\rm H.c.},
    \label{H_int}
\end{align}
where the vector 
\begin{align}
    {\pmb \chi}^{(\zeta)}_{\bf q}=\sum_{l=1}^{\cal N} {\cal J}_l\left({\pmb \xi_l^{\ast}}{\cal T}_{l,\zeta}({\bf q})+{\pmb \xi}_l {\cal T}^{\ast}_{l,\zeta+\cal N}({\bf q})\right).
    \label{chi}
\end{align}

\subsection{Two-sublattice model of AFM}

For magnons in the two-sublattice model with a lattice constant $a$ [Fig.~\ref{model}(a)] interacting with the electrons in a heavy metal of thickness $d$~\cite{s_d_Ham,interface20,interface7},
\begin{align}
    \hat{H}_{\rm int}&=\frac{a}{d}\sqrt{\frac{S}{2N_{\rm int}}}\sum_{{\bf k}{\bf k}'{\bf q}}\sum_{\zeta=1,2}\sum_{\alpha\beta}
    {\pmb \sigma}_{\alpha \beta} \cdot {\pmb \chi}^{(\zeta)}_{\bf q} \nonumber\\
    &\times\hat{c}_{{{\bf k}'},\alpha}^{\dagger}\hat{c}_{{{\bf k}},\beta}\hat{b}_{\zeta,{\bf q}}\delta_{{{\bf k}}+{\bf q},{\bf k}'}+{\rm H.c.},
    \label{H_int}
\end{align}
in which ${\pmb \sigma}=(\sigma_x,\sigma_y,\sigma_z)$ is a vector of Pauli matrices, $\hat{c}_{{\bf k},\beta}$ represents the annihilation operator of electrons, and the vectors with the superscript indicating the modes
\begin{align}
    \pmb{\chi}^{(1)}_{\bf q}&={\cal J}_1\delta {\bf S}^{(1)}_{1,{\bf q}}+{\cal J}_2\delta {\bf S}^{(1)}_{2,{\bf q}},\nonumber\\
    \pmb{\chi}^{(2)}_{\bf q}&={\cal J}_1\delta {\bf S}^{(2)}_{1,{\bf q}}+{\cal J}_2\delta {\bf S}^{(2)}_{2,{\bf q}},
    \label{chi_alpha_beta}
\end{align}
are governed by exchange couplings ${\cal J}_{1,2}$ and polarizations of sublattice spins $\delta {\bf S}_{1,{\bf q}}^{(1)}=(\sqrt{2}/2)(-B_{1,\bf q}\cos\theta\hat{\bf x}+iA_{1,\bf q}\hat{\bf y} +B_{1,\bf q}\sin\theta\hat{\bf z})$ and $\delta {\bf S}_{2,{\bf q}}^{(1)}=(\sqrt{2}/2)(-B_{1,\bf q}\cos\theta\hat{\bf x}-iA_{1,\bf q}\hat{\bf y}-B_{1,\bf q}\sin\theta\hat{\bf z})$ in mode ``$1$"
and $\delta {\bf S}_{1,{\bf q}}^{(2)}=(\sqrt{2}/2)(A_{2,{\bf q}}\cos\theta\hat{\bf x}- iB_{2,{\bf q}}\hat{\bf y}-A_{2,{\bf q}}\sin\theta\hat{\bf z})$ and $\delta {\bf S}_{2,{\bf q}}^{(2)}=(\sqrt{2}/2)(-A_{2,{\bf q}}\cos\theta\hat{\bf x}-iB_{2,{\bf q}}\hat{\bf y}-A_{2,{\bf q}}\sin\theta\hat{\bf z})$ in mode ``$2$", as shown in Fig.~\ref{model}(b) and (c).
When the exchange interaction $8JS\gg \hbar\omega_{1,2}({\bf q})$ dominates over the other energies, $A_{1,2}({\bf q})\approx \sqrt{8JS/\hbar\omega_{1,2}({\bf q})}$ and $B_{1,2}({\bf q})\approx 1/A_{1,2}({\bf q})$, and, thereby, $A_{1,2}({\bf q})\gg B_{1,2}({\bf q})$. When ${\cal J}_1={\cal J}_2$, $|\pmb{\chi}^{(1),(2)}_{\bf q}|\propto B_{1,2}({\bf q})$; when ${\cal J}_1\gg {\cal J}_2$ or ${\cal J}_2\gg {\cal J}_1$, $|\pmb{\chi}^{(1),(2)}_{\bf q}|\propto A_{1,2}({\bf q})$, so the electron-magnon coupling is enhanced when ${\cal J}_1\neq {\cal J}_2$.

$|\pmb{\sigma}_{\alpha\beta}\cdot \pmb{\chi}^{(\zeta)}_{\bf q}|^2$ is responsible for the excitation of magnon population in the $\zeta$-mode; $(\pmb{\sigma}_{\alpha\beta}\cdot \pmb{\chi}^{(1)}_{\bf q})(\pmb{\sigma}_{\beta\alpha}\cdot \pmb{\chi}^{(2)\ast}_{\bf q})$ denotes the mixing excitation of both modes in a net spin-flip process, which creates a magnon \textcolor{blue}{correlation} as in Fig.~\ref{inj_model}(b). The latter process vanishes in collinear AFMs.
Since the magnon population carries no spin along $\hat{\bf z}$, the electron spin is not conserved in the spin-flip scattering \eqref{H_int} since $\pmb{\chi}^{(1,2)}_{\bf q} \nparallel \hat{\bf z}$.
When the canted angle $\theta$ is small,  $\chi_z \rightarrow 0$ such that the  interfacial spin-conserving scattering $|\pmb{\sigma}_{\alpha\alpha}\cdot \pmb{\chi}^{(\zeta)}_{\bf q}|^2 \rightarrow 0$. Consequently, the excitation of magnon correlation is dominated by electron spin-flip scattering.

\subsection{Electrical spin injection and dephasing}

To address how magnon correlation is injected, how it relaxes, and how it transports, we establish the quantum kinetic equation for magnons. In the density matrix 
${\rho}({\bf q})=\langle\hat{\rho}_{\bf q}\rangle=\left\langle(\hat{\Psi}^{\dagger}_{\bf q})^{T}\hat{\Psi}^{T}_{\bf q}\right\rangle$ of magnons with $\langle\cdots\rangle$ denoting the ensemble average, the diagonal elements represent the magnon population, while the off-diagonal elements denote the correlation between different magnon modes. The quantum kinetic equation  $\partial_t {\rho}_{\bf q}=\partial_t {\rho}_{\bf q}|_{\rm coh}+\partial_t {\rho}_{\bf q}|_{\rm inj}+\partial_t {\rho}_{\bf q}|_{\rm rex}$ contains three contributions (refer to Appendix~\ref{kinetic_equation} for the derivation). The coherent dynamics 
\begin{align}
\partial_t {\rho}_{\bf q}|_{\rm coh}=(1/i\hbar)\left({\rho}_{\bf q} \sigma_3{\cal E}_{\bf q}-{\cal E}_{\bf q}\sigma_3{\rho}_{\bf q}\right),
\end{align}
induced by the bare Hamiltonian $\hat{H}_m=\sum_{{\bf q},\zeta}\hbar\omega_{\zeta,{\bf q}}\hat{b}_{\zeta,{\bf q}}^{\dagger}\hat{b}_{\zeta,{\bf q}}$,  originates from the intrinsic spin torque (\ref{torque}). \textcolor{blue}{The exchange of spin with metals due to $\hat{H}_{\rm int}$ is governed by 
\begin{align}
    \partial_t {\rho}_{\bf q}|_{\rm inj}&= \frac{\pi}{\hbar}\frac{S}{2N_{\rm int}}\frac{a^2}{d^2}\sum_{{\bf k},{\bf k'}}\sum_{\alpha,\beta}[f_{{\bf k'},\alpha}(1-f_{{\bf k},\beta})(\sigma_3+{\rho}_{\bf q})\nonumber\\
    &-f_{{\bf k},\beta}(1-f_{{\bf k'},\alpha})\rho_{\bf q}]{\cal M}_{\alpha \beta}({\bf k},{\bf k}')\delta_{{\bf k}+{\bf q},{\bf k}'}+{\rm H.c.},
     \label{spin_injection}
\end{align}
which may be intuitively interpreted by Fermi's Golden Rule: $f_{{\bf k'},\alpha}(1-f_{{\bf k},\beta})(\sigma_3+\rho_{{\bf q}})$ represents the magnon excitation process and $(1-f_{{\bf k'},\alpha})f_{{\bf k},\beta}\rho_{{\bf q}}$ represents the magnon absorption process. The scattering potentials are included in the $4\times4$ matrix
\begin{align}
    {\cal M}_{\alpha \beta}({\bf k},{\bf k'})=\begin{pmatrix}
        {A}_{2\times 2}({\bf k},{\bf k'}) & {C}_{2\times 2}({\bf k},{\bf k'}) \\
        {D}_{2\times 2}({\bf k},{\bf k'}) & {B}_{2\times 2}({\bf k},{\bf k'})
    \end{pmatrix}.
\end{align}
}
\begin{widetext}
\textcolor{blue}{
In the blocks  
\[{A}_{2\times 2}({\bf k},{\bf k'})= 
\begin{pmatrix}
    |{{\pmb \sigma}_{\alpha \beta}\cdot {\pmb \chi}^{(1)}_{{\bf q}}}|^2\delta(\varepsilon_{{\bf k}}-\varepsilon_{{\bf k}'}+\hbar\omega_{1,{\bf q}}) & ({{\pmb \sigma}_{\alpha \beta}\cdot {\pmb \chi}^{(1)}_{{\bf q}}})({\pmb \sigma}_{\beta \alpha}\cdot {\pmb \chi}_{{\bf q}}^{(2)\ast})\delta(\varepsilon_{{\bf k}}-\varepsilon_{{\bf k}'}+\hbar\omega_{1,{\bf q}})\\
    ({{\pmb \sigma}_{\alpha \beta}\cdot {\pmb \chi}^{(2)}_{{\bf q}}})({\pmb \sigma}_{\beta \alpha}\cdot {\pmb \chi}_{{\bf q}}^{(1)\ast})\delta(\varepsilon_{{\bf k}}-\varepsilon_{{\bf k}'}+\hbar\omega_{2,{\bf q}}) & |{{\pmb \sigma}_{\alpha \beta}\cdot {\pmb \chi}^{(2)}_{{\bf q}}}|^2\delta(\varepsilon_{{\bf k}}-\varepsilon_{{\bf k}'}+\hbar\omega_{2,{\bf q}})
\end{pmatrix}
\]
and
\[{B}_{2\times 2}({\bf k},{\bf k'})= 
\begin{pmatrix}
    -|{{\pmb \sigma}_{\alpha \beta}\cdot {\pmb \chi}_{-{\bf q}}^{(1)\ast}}|^2\delta(\varepsilon_{{\bf k}}-\varepsilon_{{\bf k}'}-\hbar\omega_{1,-{\bf q}}) & -({{\pmb \sigma}_{\alpha \beta}\cdot {\pmb \chi}_{-{\bf q}}^{(1)\ast}})({\pmb \sigma}_{\beta \alpha}\cdot {\pmb \chi}^{(2)}_{-{\bf q}})\delta(\varepsilon_{{\bf k}}-\varepsilon_{{\bf k}'}-\hbar\omega_{1,-{\bf q}})\\
    -({{\pmb \sigma}_{\alpha \beta}\cdot {\pmb \chi}_{-{\bf q}}^{(2)\ast}})({\pmb \sigma}_{\beta \alpha}\cdot {\pmb \chi}^{(1)}_{-{\bf q}})\delta(\varepsilon_{{\bf k}}-\varepsilon_{{\bf k}'}-\hbar\omega_{2,-{\bf q}}) & -|{{\pmb \sigma}_{\alpha \beta}\cdot {\pmb \chi}_{-{\bf q}}^{(2)\ast}}|^2\delta(\varepsilon_{{\bf k}}-\varepsilon_{{\bf k}'}-\hbar\omega_{2,-{\bf q}})\\
\end{pmatrix},
\]
the diagonal terms are related to the excitation of magnon population in modes ``1" and ``2", while the off-diagonal terms drive the correlation $\langle\hat{b}_{1,{\bf q}}^{\dagger}\hat{b}_{2,{\bf q}}\rangle$ and $\langle\hat{b}_{2,{\bf q}}^{\dagger}\hat{b}_{1,{\bf q}}\rangle$ between modes ``1" and ``2".
In the blocks 
\[
    {C}_{2\times 2}({\bf k},{\bf k'})= 
\begin{pmatrix}
    -({{\pmb \sigma}_{\alpha \beta}\cdot {\pmb \chi}^{(1)}_{{\bf q}}})({\pmb \sigma}_{\beta \alpha}\cdot {\pmb \chi}^{(1)}_{-{\bf q}})\delta(\varepsilon_{{\bf k}} - \varepsilon_{{\bf k}'} + \hbar\omega_{1,{\bf q}})  & -({{\pmb \sigma}_{\alpha \beta}\cdot {\pmb \chi}^{(1)}_{{\bf q}}})({\pmb \sigma}_{\beta \alpha}\cdot {\pmb \chi}^{(2)}_{-{\bf q}})\delta(\varepsilon_{{\bf k}} - \varepsilon_{{\bf k}'} + \hbar\omega_{1,{\bf q}})\\
    -({{\pmb \sigma}_{\alpha \beta}\cdot {\pmb \chi}^{(2)}_{{\bf q}}})({\pmb \sigma}_{\beta \alpha}\cdot {\pmb \chi}^{(1)}_{-{\bf q}})\delta(\varepsilon_{{\bf k}} - \varepsilon_{{\bf k}'} + \hbar\omega_{2,{\bf q}}) & -({{\pmb \sigma}_{\alpha \beta}\cdot {\pmb \chi}^{(2)}_{{\bf q}}})({\pmb \sigma}_{\beta \alpha}\cdot {\pmb \chi}^{(2)}_{-{\bf q}})\delta(\varepsilon_{{\bf k}} - \varepsilon_{{\bf k}'} + \hbar\omega_{2,{\bf q}})
    \end{pmatrix}
\]
and
\[
    {D}_{2\times 2}({\bf k},{\bf k'})= 
\begin{pmatrix}
    ({{\pmb \sigma}_{\alpha \beta}\cdot {\pmb \chi}_{-{\bf q}}^{(1)\ast}})({\pmb \sigma}_{\beta \alpha}\cdot {\pmb \chi}_{{\bf q}}^{(1)\ast})\delta(\varepsilon_{{\bf k}}-\varepsilon_{{\bf k}'}-\hbar\omega_{1,-{\bf q}}) & ({{\pmb \sigma}_{\alpha \beta}\cdot {\pmb \chi}_{-{\bf q}}^{(1)\ast}})({\pmb \sigma}_{\beta \alpha}\cdot {\pmb \chi}_{{\bf q}}^{(2)\ast})\delta(\varepsilon_{{\bf k}}-\varepsilon_{{\bf k}'}-\hbar\omega_{1,-{\bf q}})\\
    ({{\pmb \sigma}_{\alpha \beta}\cdot {\pmb \chi}_{-{\bf q}}^{(2)\ast}})({\pmb \sigma}_{\beta \alpha}\cdot {\pmb \chi}_{{\bf q}}^{(1)\ast})\delta(\varepsilon_{{\bf k}}-\varepsilon_{{\bf k}'}-\hbar\omega_{2,-{\bf q}}) & ({{\pmb \sigma}_{\alpha \beta}\cdot {\pmb \chi}_{-{\bf q}}^{(2)\ast}})({\pmb \sigma}_{\beta \alpha}\cdot {\pmb \chi}_{{\bf q}}^{(2)\ast})\delta(\varepsilon_{{\bf k}}-\varepsilon_{{\bf k}'}-\hbar\omega_{2,-{\bf q}})
\end{pmatrix},
\]
the diagonal terms govern the generation of intraband magnon pairs $\langle \hat{b}^{\dagger}_{\xi,{\bf q}}\hat{b}^{\dagger}_{\xi,-{\bf q}}\rangle$ and $\langle \hat{b}_{\xi,{\bf q}}\hat{b}_{\xi,-{\bf q}}\rangle$, while the off-diagonal terms drive the interband pairs $\langle \hat{b}^{\dagger}_{\xi,{\bf q}}\hat{b}^{\dagger}_{\xi',-{\bf q}}\rangle$ and $\langle \hat{b}_{\xi,{\bf q}}\hat{b}_{\xi',-{\bf q}}\rangle$.}
\end{widetext}
The electron population $f(\varepsilon_{{\bf k},\beta})=f_0(\varepsilon_{{\bf k}}-\varepsilon_{F})-\delta \mu_{\beta}{\partial f_0}/{\partial \varepsilon_{{\bf k}}}$ deviates from the equilibrium 
$f_0(\varepsilon_{{\bf k}}-\varepsilon_{F})=\{{\rm exp}[(\varepsilon_{{\bf k}}-\varepsilon_{F})/(k_BT)]+1\}^{-1}$ by a spin accumulation $\delta\mu_{\beta}$ that is much smaller than the Fermi energy $ \varepsilon_F$, which exchanges the electron spin with the magnon population and correlation. Magnon population and correlation relax to their equilibrium $\rho_{\bf q}^{(0)}$ due to the magnon-phonon interaction within the time $\tau_m$~\cite{YFeO,tau3} via 
\begin{align}
    \partial_t {\rho}_{\bf q}|_{\rm rex}=-(\rho_{\bf q}-\rho_{\bf q}^{(0)})/\tau_m.
\end{align}
The steady-state solution is obtained by setting $\partial_t {\rho}_{\bf q}=0$. We solve the quantum kinetic equation in the linear response regime, see Appendix~\ref{linear_response_regime}.

\textcolor{blue}{We have established a complete form of the quantum kinetic equation by considering the evolution of a full density matrix. Tang \textit{et al.} recently considered the quantum kinetic equation for a subspace of the density matrix~\cite{Ping_Tang}, in which the Hanle effect is the primary focus. Besides Hanle effect, our work focuses on analyzing the microscopic origins of decoherence in the spin injection and transport: 1) Intrinsic decoherence due to the free induction decay resulting from the superposition of different oscillation frequencies; 2) Extrinsic decoherence due to spin exchange with electrons in adjacent metals, based on which we propose to gate the coherent spin transport by a metallic gate. }

\begin{figure}[htp!]	
\includegraphics[width=0.49\textwidth,trim=0.6cm 0cm 0cm 0.1cm]{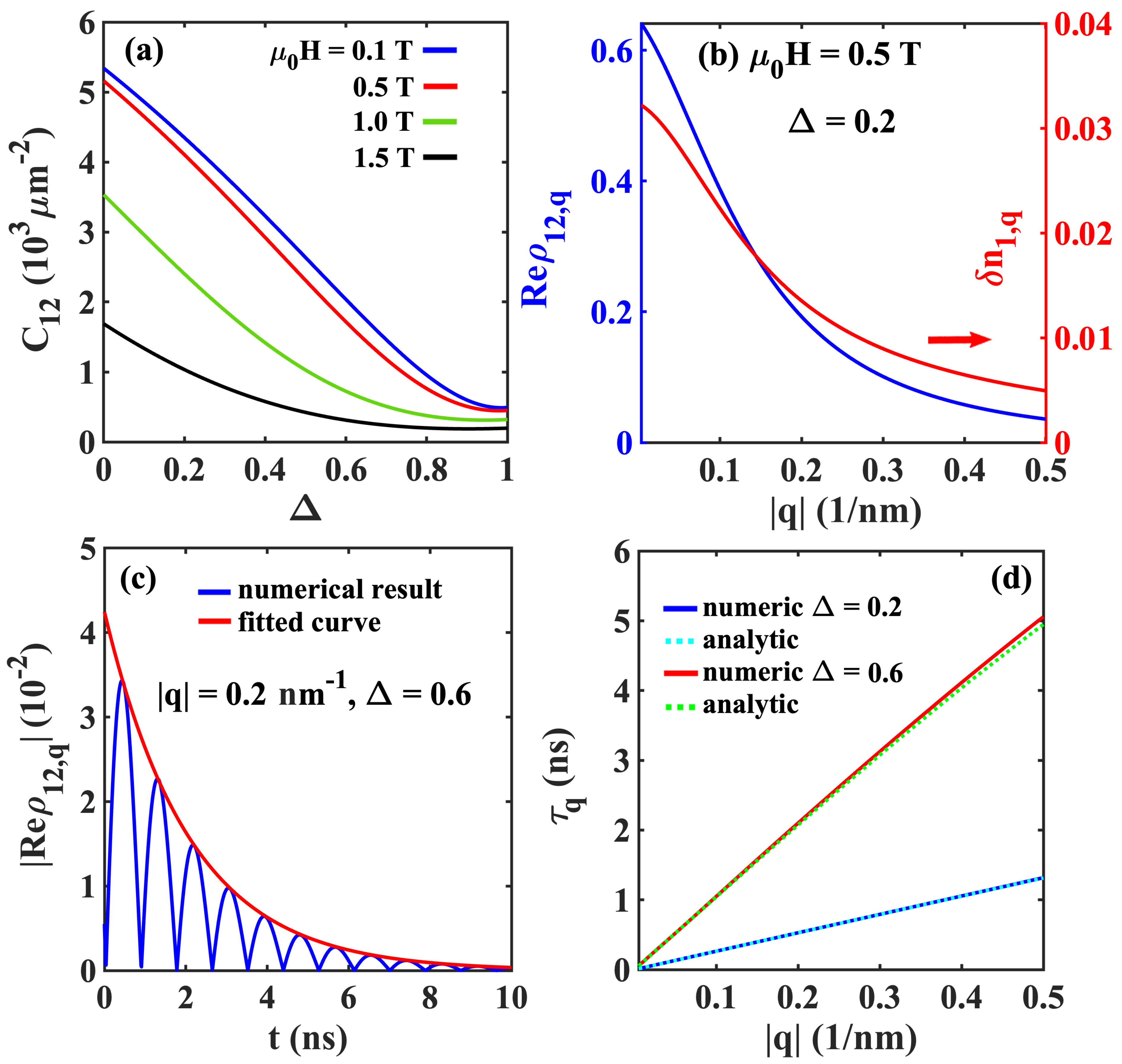}
\caption{(a) Exchange asymmetry $\Delta$ and magnetic-field dependencies of injected magnon correlation density. (b) Distribution of the injected magnon correlation and population of mode ``1" in the Brillouin zone. \textcolor{blue}{(c) Time evolution of magnon correlation ${\rm Re}\rho_{12,{\bf q}}$ (the blue curve) and the fitting via the exponential decay $\sim e^{-t/\tau_{\bf q}}$ (the red curve).} (d) Calculated dephasing time $\tau_{\bf q}$ due to the magnon-electron interaction, showing excellent agreement between numerical and analytical calculations.}
\label{inj}
\end{figure}

Figure~\ref{inj}(a) plots
the real part of magnon correlation density $C_{12}=(1/S_0){\rm Re}\sum_{\bf q}\rho_{12}({\bf q})$  injected in an area $S_0$ by the spin accumulation $\mu_s=\mu_{\uparrow}-\mu_{\downarrow}=0.1$~meV~\cite{mu1,mu2,mu3,mu4} in heavy metals with a lattice constant $a=0.5$~nm and thickness $d=100$~nm~\cite{d1} at room temperature $T=300$~K. According to Refs.~\cite{YFeO,tau3}, we take  $\tau_m\approx 1$~ns. The interfacial exchange interactions ${\cal J}_2={\cal J}$ and ${\cal J}_1={\cal J}\Delta$ can be asymmetric with $\Delta\in [0,1]$, for which we take ${\cal J}=-15$~meV~\cite{interface20,interface7,J1,J2,J3}.  As resolved in the Brillouin zone in Fig.~\ref{inj}(b), the injected magnon correlation is much larger than the population. Therefore, the magnon population that carries no spin along the N\'eel vector is only weakly excited owing to spin non-conservation at the interface.

The spin density \textcolor{blue}{$S_z\approx (1/S_0)2{\rm Re}\sum_{\bf q}\rho_{12}({\bf q})$} along the N\'eel vector is solely carried by magnon correlation. The net spin injection is suppressed by decreasing the asymmetry $\Delta$ and increasing the magnetic field $H$.
This can be attributed to the spin dephasing due to the transfer of magnon spin to the conductor and the intrinsic torque exerted on magnon correlation, which restricts spin injection, as analyzed below.

Focusing on spin dephasing due to magnon-electron coupling, we set $\mu_{\uparrow}=\mu_{\downarrow}$ in Eq.~\eqref{spin_injection}. Non-equilibrium magnon correlations now transfer spin angular momentum back to electrons. 
\textcolor{blue}{The dephasing time $\tau_{\bf q}$ in the numerical result is obtained by the fitted evolution of the magnon correlation ${\rm Re}\rho_{12,{\bf q}}(t)$ by the exponential function $\sim e^{-t/\tau_{\bf q}}$, as shown in Fig.~\ref{inj}(c).
In the numerical calculation, the overall additional relaxation $\tau_m$ is disregarded, such that the decay originates only from the dephasing. Analytically,}
disregarding two-magnon excitations $\langle \hat{b}^{\dagger}_{\bf q} \hat{b}^{\dagger}_{-\bf q}\rangle$, the magnon correlation $\rho_{12}({\bf q})$ obeys
\begin{align}
    {\partial \rho_{12}({\bf q})}/{\partial t}=i\delta\omega_{\bf q}\rho_{12}({\bf q})-\left({1}/{\tau_{\bf q}}+{1}/{\tau_m}\right)\rho_{12}({\bf q}).
\end{align}
The frequency difference $|\delta\omega_{\bf q}|=|\omega_{1,{\bf q}}-\omega_{2,{\bf q}}|$ that is increased by applied magnetic fields [inset in Fig.~\ref{model}(d)] causes the oscillation of the correlation. It causes free-induction decay of $S_z$ due to the interference of different oscillation frequencies, which is enhanced by the magnetic field. On the other hand, 
\begin{align}
    \frac{1}{\tau_{\bf q}}=\frac{1}{\hbar}\frac{Sa^2}{2d^2}C_{q}\sum_{\alpha\beta}\sum_{\zeta=1,2}\hbar\omega_{\zeta,{\bf q}}|\pmb{\sigma}_{\alpha\beta}\cdot \pmb{\chi}^{(\zeta)}_{\bf q}|^2
    \label{tau_q}
\end{align}
is responsible for the pure decay of correlations. Here, $C_q=a^2dm^2/(4\pi\hbar^4q)$; the decay rate $1/\tau_{\bf q}$ does not depend on the 
magnetic field but is suppressed by increasing the exchange asymmetry $\Delta$ as in Fig.~\ref{inj}(d). This is because according to Eq.~\eqref{chi_alpha_beta}, the coupling constant
$\sum_{\alpha\beta}\hbar\omega_{\zeta}({\bf q})|\pmb{\sigma}_{\alpha\beta}\cdot \pmb{\chi}^{(\zeta)}_{\bf q}|^2 \sim 8JS{\cal J}^2(1-\Delta)^2$
is suppressed by increasing $\Delta\in [0,1]$. The net dephasing of $S_z$ is governed by the joint effect of free-induction decay $\delta \omega_{\bf q}$ and the decay rates $1/\tau_{\bf q}$ and $1/\tau_m$.

\section{Gated nonlocal spin transport}

\label{magnon_transport}

\begin{figure}[htp!]	
\includegraphics[width=0.48\textwidth,trim=0.6cm 0cm 0cm 0.1cm]{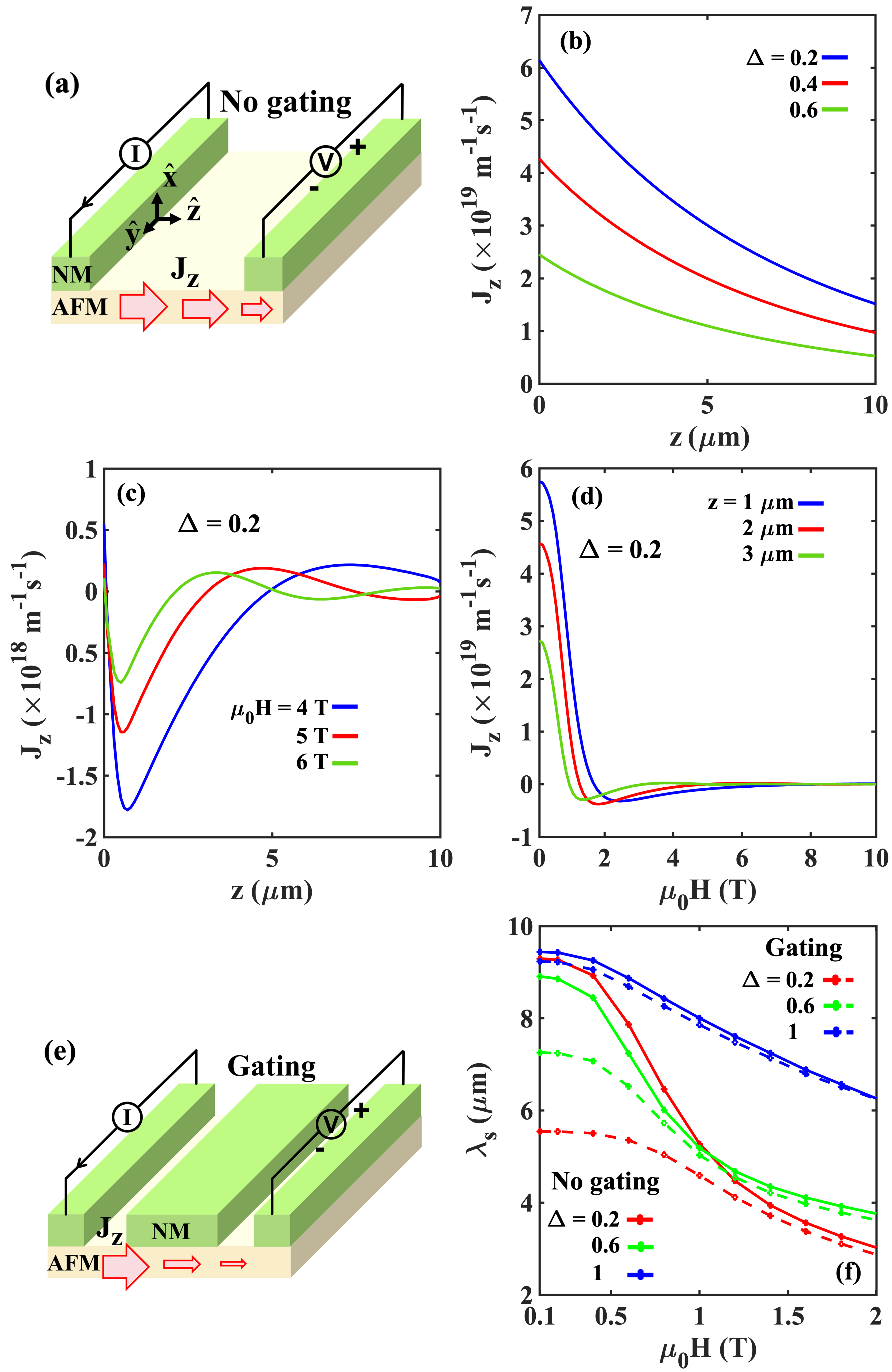}
\caption{Nonlocal spin transport tunable by the magnetic fields and metallic gate. 
(a) Schematic of a nonlocal spin transport device. (b) Diffusive spin current density $J_z(z)$ along the $\hat{\bf z}$-direction. 
(c) Hanle oscillations at high magnetic field. \textcolor{blue}{(d) Spin current of fixed position $z$ as a function of the magnetic field as a signature of the Hanle effect.} (e) Gated spin transport by normal metals. (f) Spin diffusion length with and without metallic gates, also showing magnetic-field tunability.}
\label{spin_transport}
\end{figure}

The injected magnon correlations carry net spins along the N\'eel vector and diffuse away from the injector located around $z=0$, resulting in a spin current as depicted in Fig.~\ref{spin_transport}(a). \textcolor{blue}{For the density matrix, the diffusion term reads
\begin{align}
    \partial_t \rho_{\bf q}|_{\rm diff}=(\pmb{\cal V}_{\bf q}\sigma_3\cdot\nabla_{\bf r}\rho_{\bf q}+\nabla_{\bf r}\rho_{\bf q}\cdot\sigma_3\pmb{\cal V}_{\bf q})/2.
\end{align}} 
Accordingly, in the steady state, the diffusion along the $\hat{\bf z}$-direction away from the injector is governed by
\begin{align}
    -v_z({\bf q})\frac{\partial\rho_{12}(z,{\bf q})}{\partial z}-\left(\frac{1}{\tau_m}-i\delta\omega_{\bf q}\right)\rho_{12}(z,{\bf q})=0,
    \label{diff1}
\end{align}
where $v_z({\bf q})=(v_{1,z}({\bf q})+v_{2,z}({\bf q}))/2$ is the group velocity of the magnon correlation along the $\hat{\bf z}$-direction and $\delta\omega_{\bf q}=\omega_{1,{\bf q}}-\omega_{2,{\bf q}}$ is the frequency difference of the two magnon modes. 
By substituting $\rho_{12}(z,{\bf q})=-v_z\tau_{\rm eff}({\bf q})\partial_z\rho_{12}(z,{\bf q})$ back to Eq.~(\ref{diff1}) with 
an effective dephasing rate $1/\tau_{\rm eff}({\bf q})=1/\tau_m-i\delta\omega_{\bf q}$, we arrive at the second-order diffusion equation 
\begin{align}
    D_{\bf q}\partial^2_z\rho_{12}(z,{\bf q})-{\rho_{12}(z,{\bf q})}/{\tau_{\rm eff}({\bf q})}=0,    
    \label{diffusion}
\end{align}
which is governed by the diffusion coefficient $D_{\bf q}=v_z^2({\bf q})\tau_{\rm eff}({\bf q})$ and  the effective dephasing rate $1/\tau_{\rm eff}({\bf q})=1/\tau_m-i\delta\omega_{\bf q}$, with $v_z({\bf q})=(v_{1,z}({\bf q})+v_{2,z}({\bf q}))/2$ being the group velocity of magnon correlation. 
Equation~(\ref{diffusion}) has the general solution
\begin{align}
    \rho_{12}(z>0,{\bf q})&=\alpha_{\bf q}\exp\left(-\frac{z}{|v_z({\bf q})|\tau_{\rm eff}({\bf q})}\right),\nonumber\\
    \rho_{12}(z<0,{\bf q})&=\beta_{\bf q}\exp\left(\frac{z}{|v_z({\bf q})|\tau_{\rm eff}({\bf q})}\right). 
\end{align}
At $z=0$, the continuity requires the coefficients $\alpha_{\bf q}=\beta_{\bf q}=\rho^{(0)}_{12}({\bf q})$, where $\rho^{(0)}_{12}({\bf q})$ is given by the steady-state injected magnon correlation beneath the injector.

With the correlation accumulation $\rho^{(0)}_{12}({\bf q})$ beneath the injector and according to Eq.~(\ref{Jsz}), Eq.~(\ref{diffusion}) leads to the spin-current density in the region $z>0$
\begin{align}
    J_z(z)&=\frac{2}{S_0}\sum_{\bf q}|v_{z,{\bf q}}|{\cal A}_{\bf q}{\rm Re}\left[\rho^{(0)}_{12}({\bf q})\exp(-\frac{z}{|v_{z,{\bf q}}|\tau_{\rm eff}({\bf q})})\right].
    \nonumber
\end{align}

Figure~\ref{spin_transport}(b) plots the exponential decay of spin-current densities $J_z(z)$ along the transport $\hat{\bf z}$-direction at a small magnetic field $\mu_0H=0.5$~T;
when the magnetic field is sufficiently strong, the spin current becomes oscillating as shown in Fig.~\ref{spin_transport}(c) since in this case  $|\delta\omega_{\bf q}| \gg 1/\tau_m$, such that $J_z(z)$ shows a pronounced Hanle oscillation. 
This is consistent with experimental observations~\cite{easy_plane6}, \textcolor{blue}{which discusses the Hanle effect in \textit{easy-plane} antiferromagnets, with a peak signal arising at a large magnetic field $\sim 8$~T. We predict the Hanle effect in \textit{canted easy-axis} antiferromagnets, where the spin-current peak appears at low magnetic fields, as shown in Fig.~\ref{spin_transport}(d) for fixed position. The physical origin of the Hanle effect in both descriptions is essentially the same. In the former pseudospin description~\cite{easy_plane6,easy_plane8}, this is interpreted as the precession of a pseudospin around a pseudo-magnetic field that is governed by the frequency difference between two modes. In our quantum description, the magnon correlation oscillates in the frequency difference between the two magnon modes. The pseudospin description employs an approximation by using an ensemble-averaged frequency difference, whereas the quantum description applies no approximation by using a full numerical calculation.}

As addressed by Eq.~\eqref{tau_q}, the passive metallic gate is a source of spin dephasing. To detect such spin dephasing, we propose to exploit the normal metals to gate the nonlocal spin transport~\cite{easy_axis13} in a device illustrated in Fig.~\ref{spin_transport}(e). The random transfer of spin between normal metals and AFM brakes the spin flow by decreasing the spin diffusion length $\lambda_s$ at low magnetic fields through the modified effective dephasing rate $1/\tau_{\rm eff}({\bf q})\rightarrow 1/\tau_m+1/\tau_{\bf q}-i\delta\omega_{\bf q}$, as shown in Fig.~\ref{spin_transport}(f). Nevertheless, at higher magnetic fields, the free-induction decay $\delta\omega_{\bf q}$ dominates the spin transport, so the spin flow becomes immune to the metallic gate. \textcolor{blue}{The efficiency of gating strongly depends on the asymmetry $\Delta$ of the coupling of the sublattice magnetic moments to electrons $\propto (1-\Delta)^2$. Here, $\Delta=1$ corresponds to a perfectly compensated interface, while $\Delta=0$ corresponds to an uncompensated interface. Defective interfaces enhance electron-magnon scattering, resulting in a stronger gating efficiency.}

\section{Discussion and conclusion}

\label{summary}

The build-up of observable quantum coherence between magnon modes and their transport in thermodynamics is surprising in that no definite phase difference between two modes is promised. This is rooted in a mixing
excitation of both modes in a quantum spin-flip process, which creates a magnon \textcolor{blue}{correlation}. The injected magnon correlation oscillates with the frequency difference of two modes, which interfere with each other for different wave vectors. The adjacent metal plays a twofold role: it injects magnon correlation but also causes dephasing of magnon correlation due to their backaction on electron spin dynamics. Our quantum theory establishes a comprehensive understanding of phase-coherent \textcolor{blue}{magnon correlation} formation in canted AFMs, predicts a practical electrical scheme for manipulating coherent spin signals in thermal magnon transport, and applies to  $\alpha$-Fe$_2$O$_3$~\cite{easy_plane5,easy_plane6,easy_plane7}, FePS$_3$~\cite{FePS}, Cr$_2$O$_3$~\cite{easy_axis1,CrO}, and many other noncollinear AFMs.

\begin{acknowledgments}
This work is financially supported by the National Key Research and Development Program of China under Grant No.~2023YFA1406600 and the National Natural Science Foundation of China under Grant No.~12374109. 
\end{acknowledgments}

\begin{appendix}

\section{Quantum kinetic equation}

\label{kinetic_equation}

We derive the quantum kinetic equation for magnons to analyze the spin injection and spin dephasing processes, which can conveniently resolve which magnon population and correlation, and which wave vectors, are injected, and how their non-equilibrium spin relaxes and transports.

In the density operator of magnons
\[
\hat{\rho}_{\bf q}=\begin{pmatrix}
        \hat{b}_{1,{\bf q}}^{\dagger}\hat{b}_{1,{\bf q}} & \hat{b}_{1,{\bf q}}^{\dagger}\hat{b}_{2,{\bf q}}&
        \hat{b}_{1,{\bf q}}^{\dagger}\hat{b}_{1,-{\bf q}}^{\dagger} & \hat{b}_{1,{\bf q}}^{\dagger}\hat{b}_{2,-{\bf q}}^{\dagger}\\
        \hat{b}_{2,{\bf q}}^{\dagger}\hat{b}_{1,{\bf q}} & \hat{b}_{2,{\bf q}}^{\dagger}\hat{b}_{2,{\bf q}}&
        \hat{b}_{2,{\bf q}}^{\dagger}\hat{b}_{1,-{\bf q}}^{\dagger} & \hat{b}_{2,{\bf q}}^{\dagger}\hat{b}_{2,-{\bf q}}^{\dagger}\\
        \hat{b}_{1,-{\bf q}}\hat{b}_{1,{\bf q}} & \hat{b}_{1,-{\bf q}}\hat{b}_{2,{\bf q}} &
        \hat{b}_{1,-{\bf q}}\hat{b}_{1,-{\bf q}}^{\dagger} & \hat{b}_{1,-{\bf q}}\hat{b}_{2,-{\bf q}}^{\dagger}\\
        \hat{b}_{2,-{\bf q}}\hat{b}_{1,{\bf q}} & \hat{b}_{2,-{\bf q}}\hat{b}_{2,{\bf q}} &
        \hat{b}_{2,-{\bf q}}\hat{b}_{1,-{\bf q}}^{\dagger} & \hat{b}_{2,-{\bf q}}\hat{b}_{2,-{\bf q}}^{\dagger}
    \end{pmatrix},
\] 
the diagonal elements represent the magnon population, while the off-diagonal elements denote the correlations between different magnon modes.
With the total Hamiltonian  $\hat{H}(t) = \hat{H}_m + \hat{H}_{\rm int}^{I}(t)$ in the interaction representation ``$I$", the ensemble average of $\hat{\rho}_{\bf q}$ evolves according to~\cite{Mahan} 
\begin{align}
   i\hbar \frac{\partial {\rho}_{\bf q}}{\partial t}&= i\hbar \frac{\partial \langle \hat{\rho}_{\bf q}\rangle}{\partial t}=i\hbar \frac{\partial}{\partial t}{\rm Tr}\left(\hat{\rho}_I(t)\hat{\rho}_{\bf q}^I(t)\right) \nonumber\\
    &={\rm Tr}\left[\hat{\rho}_I(t)\left(i\hbar \frac{\partial}{\partial t}\hat{\rho}_{\bf q}^I(t)\right)+\left(i\hbar \frac{\partial}{\partial t}\hat{\rho}_I(t)\right)\hat{\rho}_{\bf q}^I(t)\right] \nonumber\\
    &=i\hbar \partial_t \langle \hat{\rho}_{\bf q}\rangle|_{\rm coh}+i\hbar \partial_t \langle \hat{\rho}_{\bf q}\rangle|_{\rm inj},
    \label{A1}
\end{align}
in which $\langle\cdots\rangle$ represents the ensemble average and $\hat{\rho}_I$ is the density matrix of magnons. Here, $i\hbar \partial_t \langle \hat{\rho}_{\bf q}\rangle|_{\rm coh}={\rm Tr}\left(\hat{\rho}_I(t)[\hat{\rho}_{\bf q}^I(t),\hat{H}_m]\right)$ represents the coherent dynamics induced by the intrinsic Hamiltonian $\hat{H}_m$ and $i\hbar \partial_t \langle \hat{\rho}_{\bf q}\rangle|_{\rm inj}={\rm Tr}\left(\hat{\rho}_I(t)[\hat{\rho}_{\bf q}^I(t),\hat{H}_{\rm int}^I(t)]\right)$ represents the dynamics caused by the external environment $\hat{H}_{\rm int}$, \textit{e.g.}, the proximity conductors.

\begin{widetext}
The density matrix $\hat{\rho}_I$ is governed by the Liouville equation $i\hbar\partial \hat{\rho}_I(t)/\partial t=[\hat{H}_{\rm int}^I(t),\hat{\rho}_I(t)]$. 
Its formal solution
$\hat{\rho}_I(t)=\hat{\rho}_0+({1}/{i\hbar})\int_{-\infty}^{t}[\hat{H}_{\rm int}^I(t'),\hat{\rho}_I(t')]dt'$,
where $\hat{\rho}_0=\hat{\rho}_I(t=-\infty)$ is the magnon density matrix at equilibrium without introducing interaction. 
Accordingly, 
\[
i\hbar\partial_t \langle \hat{\rho}_{\bf q}\rangle|_{\rm inj}
    =\frac{1}{i\hbar}\int_{-\infty}^t {\rm Tr}\left(\hat{\rho}_I(t')\left[[\hat{\rho}_{\bf q}^I(t),\hat{H}_{\rm int}^I(t)],\hat{H}_{\rm int}^I(t')\right]\right)dt',
\]
noting ${\rm Tr}\left(\hat{\rho}_0 [\hat{\rho}_{\bf q}^I(t),\hat{H}_{\rm int}^I(t)] \right)={\rm Tr}\left( \hat{\rho}_{\bf q}^I(t)[\hat{H}_{\rm int}^I(t),\hat{\rho}_0] \right)$ vanishes. 
When the interaction is weak, we may apply the Born-Markov approximation~\cite{Open_Quantum_Systems} with $\hat{\rho}_I(t')=\hat{\rho}_I(t)$ and arrive at 
\begin{align}
    \partial_t \langle \hat{\rho}_{\bf q}\rangle|_{\rm inj}
    &=\frac{1}{i\hbar}\int_{-\infty}^t {\rm Tr}\left(\hat{\rho}_I(t)\left[\frac{1}{i\hbar}[\hat{\rho}_{\bf q}^I(t),\hat{H}_{\rm int}^I(t)],\hat{H}_{\rm int}^I(t')\right]\right)dt'.
    \label{A2}
\end{align}
With the magnon-electron interaction $\hat{H}_{\rm int}$ in Eq.~(\ref{H_int}), the commutator 
\begin{align}
     \frac{1}{i\hbar}\left[\hat{\rho}_{\bf q},\hat{H}_{\rm int}\right]
     =\frac{1}{i\hbar}\frac{a}{d}\sqrt{\frac{S}{2N_{\rm int}}}\sum_{{\bf k},{\bf k}'}\sum_{\alpha \beta}
    \begin{pmatrix}
        -{\pmb \sigma}_{\alpha \beta}\cdot{\pmb \chi}^{(1)}_{{\bf q}}\hat{c}_{{\bf k}',\alpha}^{\dagger}\hat{c}_{{\bf k},\beta}\delta_{+} \\
        -{\pmb \sigma}_{\alpha \beta}\cdot{\pmb \chi}^{(2)}_{{\bf q}}\hat{c}_{{\bf k}',\alpha}^{\dagger}\hat{c}_{{\bf k},\beta}\delta_{+} \\
        {\pmb \sigma}_{\beta \alpha}\cdot{\pmb \chi}_{-{\bf q}}^{(1)\ast}\hat{c}_{{\bf k},\beta}^{\dagger}\hat{c}_{{\bf k}',\alpha}\delta_{-} \\
        {\pmb \sigma}_{\beta \alpha}\cdot{\pmb \chi}_{-{\bf q}}^{(2)\ast}\hat{c}_{{\bf k},\beta}^{\dagger}\hat{c}_{{\bf k}',\alpha}\delta_{-}
    \end{pmatrix}
    \begin{pmatrix}
        \hat{b}_{1,{\bf q}}, & \hat{b}_{2,{\bf q}}, & \hat{b}_{1,-{\bf q}}^{\dagger}, & \hat{b}_{2,-{\bf q}}^{\dagger}
    \end{pmatrix} +{\rm H.c.},
    \label{NH1}
\end{align}
where $\delta_{\pm}=\delta_{{\bf k}\pm {\bf q},{\bf k}'}$.
In the interaction representation,
\begin{align}
    &\frac{1}{i\hbar}\left\langle \left[ \frac{1}{i\hbar}[\hat{\rho}_{\bf q}^I(t),\hat{H}_{\rm int}^I(t)],\hat{H}_{\rm int}^I(t') \right]\right\rangle \nonumber\\
    &=\left(\frac{1}{i\hbar}\right)^2\frac{a}{d}\sqrt{\frac{S}{2N_{\rm int}}}
    \sum_{{\bf k},{\bf k}'}\sum_{\alpha \beta} \left\{
    \left\langle \begin{pmatrix}
    -{\pmb \sigma}_{\alpha \beta}\cdot{\pmb \chi}^{(1)}_{{\bf q}}\delta_{+}[\hat{c}_{{\bf k}',\alpha}^{\dagger}(t)\hat{c}_{{\bf k},\beta}(t),\hat{H}_{\rm int}^I(t')] \\
    -{\pmb \sigma}_{\alpha \beta}\cdot{\pmb \chi}^{(2)}_{{\bf q}}\delta_{+}[\hat{c}_{{\bf k}',\alpha}^{\dagger}(t)\hat{c}_{{\bf k},\beta}(t),\hat{H}_{\rm int}^I(t')] \\
    {\pmb \sigma}_{\beta \alpha}\cdot{\pmb \chi}_{-{\bf q}}^{(1)\ast}\delta_{-}[\hat{c}_{{\bf k},\beta}^{\dagger}(t)\hat{c}_{{\bf k}',\alpha}(t),\hat{H}_{\rm int}^I(t')] \\
    {\pmb \sigma}_{\beta \alpha}\cdot{\pmb \chi}_{-{\bf q}}^{(2)\ast}\delta_{-}[\hat{c}_{{\bf k},\beta}^{\dagger}(t)\hat{c}_{{\bf k}',\alpha}(t),\hat{H}_{\rm int}^I(t')]
    \end{pmatrix}\right.
    \begin{pmatrix}
        \hat{b}_{1,{\bf q}}(t) \\ 
        \hat{b}_{2,{\bf q}}(t) \\ \hat{b}_{1,-{\bf q}}^{\dagger}(t)\\ \hat{b}_{2,-{\bf q}}^{\dagger}(t)
    \end{pmatrix}^T \right\rangle \nonumber\\
    &\left.+\left\langle \begin{pmatrix}
        -{\pmb \sigma}_{\alpha \beta}\cdot{\pmb \chi}^{(1)}_{{\bf q}}\delta_{+}\hat{c}_{{\bf k}',\alpha}^{\dagger}(t)\hat{c}_{{\bf k},\beta}(t) \\
        -{\pmb \sigma}_{\alpha \beta}\cdot{\pmb \chi}^{(2)}_{{\bf q}}\delta_{+}\hat{c}_{{\bf k}',\alpha}^{\dagger}(t)\hat{c}_{{\bf k},\beta}(t) \\
        {\pmb \sigma}_{\beta \alpha}\cdot{\pmb \chi}_{-{\bf q}}^{(1)\ast}\delta_{-}\hat{c}_{{\bf k},\beta}^{\dagger}(t)\hat{c}_{{\bf k}',\alpha}(t) \\
        {\pmb \sigma}_{\beta \alpha}\cdot{\pmb \chi}_{-{\bf q}}^{(2)\ast}\delta_{-}\hat{c}_{{\bf k},\beta}^{\dagger}(t)\hat{c}_{{\bf k}',\alpha}(t)
    \end{pmatrix}  
    \begin{pmatrix}
        [\hat{b}_{1,{\bf q}}(t),\hat{H}_{\rm int}^I(t')]\\
        [\hat{b}_{2,{\bf q}}(t),\hat{H}_{\rm int}^I(t')]\\
        [\hat{b}_{1,-{\bf q}}^{\dagger}(t),\hat{H}_{\rm int}^I(t')]\\ [\hat{b}_{2,-{\bf q}}^{\dagger}(t),\hat{H}_{\rm int}^I(t')]
    \end{pmatrix}^T \right\rangle \right\}+{\rm H.c.}. 
\end{align}
\end{widetext}
We focus on the case with the spin polarization of electrons along $\hat{\bf z}$, governed by the spin-dependent distribution $\langle\hat{c}^{\dagger}_{{\bf k}\alpha}\hat{c}_{{\bf k}'\beta}\rangle=f_{{\bf k},\alpha}\delta_{\alpha\beta}\delta_{{\bf k}{\bf k}'}$. 
With $\hat{b}_{\zeta,{\bf q}}(t)=\hat{b}_{\zeta,{\bf q}}e^{-i\omega_{\zeta,{\bf q}}t}$ and $\hat{c}_{{\bf k},\beta}(t)=\hat{c}_{{\bf k},\beta}e^{-i\varepsilon_{{\bf k}}t/\hbar}$, the magnon-electron interaction contributes to the quantum kinetic equation by 
\begin{align}
    \partial_t \langle \hat{\rho}_{\bf q}\rangle|_{\rm inj}
   &=\frac{i}{\hbar}\frac{S}{2N_{\rm int}}\frac{a^2}{d^2}\sum_{{\bf k},{\bf k'}}\sum_{\alpha,\beta}[f_{{\bf k'},\alpha}(1-f_{{\bf k},\beta})(\sigma_3+{\rho}_{\bf q}) \nonumber\\
   &-f_{{\bf k},\beta}(1-f_{{\bf k'},\alpha}){\rho}_{\bf q}]{\cal G}_{\alpha \beta}({\bf k},{\bf k}')\delta_{{\bf k}+{\bf q},{\bf k}'}+{\rm H.c.},
    \label{dNdt}
\end{align}
where the matrix 
\begin{align}
    &{\cal G}_{\alpha \beta}({\bf k},{\bf k}')=\begin{pmatrix}
        \frac{{\pmb \sigma}_{\alpha \beta}\cdot {\pmb \chi}^{(1)}_{{\bf q}}}{\varepsilon_{{\bf k}}-\varepsilon_{{\bf k}'}+\hbar\omega_{1,{\bf q}}+i0^+} \\
        \frac{{\pmb \sigma}_{\alpha \beta}\cdot {\pmb \chi}^{(2)}_{{\bf q}}}{\varepsilon_{{\bf k}}-\varepsilon_{{\bf k}'}+\hbar\omega_{2,{\bf q}}+i0^+} \\
        \frac{{\pmb \sigma}_{\alpha \beta}\cdot {\pmb \chi}_{-{\bf q}}^{(1)\ast}}{\varepsilon_{{\bf k}}-\varepsilon_{{\bf k}'}-\hbar\omega_{1,-{\bf q}}+i0^+} \\
        \frac{{\pmb \sigma}_{\alpha \beta}\cdot {\pmb \chi}_{-{\bf q}}^{(2)\ast}}{\varepsilon_{{\bf k}}-\varepsilon_{{\bf k}'}-\hbar\omega_{2,-{\bf q}}+i0^+}
    \end{pmatrix}\nonumber\\
    &\times
    \begin{pmatrix}
        {\pmb \sigma}_{\beta \alpha}\cdot {\pmb \chi}_{{\bf q}}^{(1)\ast}, & 
        {\pmb \sigma}_{\beta \alpha}\cdot {\pmb \chi}_{{\bf q}}^{(2)\ast}, &
        -{\pmb \sigma}_{\beta \alpha}\cdot {\pmb \chi}^{(1)}_{-{\bf q}}, &
        -{\pmb \sigma}_{\beta \alpha}\cdot {\pmb \chi}^{(2)}_{-{\bf q}}
    \end{pmatrix}. \nonumber
\end{align}
Disregarding the principal value of ${\cal G}_{\alpha \beta}$ that only shifts the magnon frequency, the matrix 
\begin{align}
    &{\cal G}_{\alpha \beta}({\bf k},{\bf k'})\rightarrow
    -i\pi\begin{pmatrix}
        {{\pmb \sigma}_{\alpha \beta}\cdot {\pmb \chi}^{(1)}_{{\bf q}}}\delta(\varepsilon_{{\bf k}}-\varepsilon_{{\bf k}'}+\hbar\omega_{1,{\bf q}}) \\
        {{\pmb \sigma}_{\alpha \beta}\cdot {\pmb \chi}^{(2)}_{{\bf q}}}\delta(\varepsilon_{{\bf k}}-\varepsilon_{{\bf k}'}+\hbar\omega_{2,{\bf q}}) \\
        {{\pmb \sigma}_{\alpha \beta}\cdot {\pmb \chi}_{-{\bf q}}^{(1)\ast}}\delta(\varepsilon_{{\bf k}}-\varepsilon_{{\bf k}'}-\hbar\omega_{1,-{\bf q}}) \\
        {{\pmb \sigma}_{\alpha \beta}\cdot {\pmb \chi}_{-{\bf q}}^{(2)\ast}}\delta(\varepsilon_{{\bf k}}-\varepsilon_{{\bf k'}}-\hbar\omega_{2,-{\bf q}})
    \end{pmatrix}\nonumber\\
    &\times
    \begin{pmatrix}
        {\pmb \sigma}_{\beta \alpha}\cdot {\pmb \chi}_{{\bf q}}^{(1)\ast}, & 
        {\pmb \sigma}_{\beta \alpha}\cdot {\pmb \chi}_{{\bf q}}^{(2)\ast}, &
        -{\pmb \sigma}_{\beta \alpha}\cdot {\pmb \chi}^{(1)}_{-{\bf q}}, &
        -{\pmb \sigma}_{\beta \alpha}\cdot {\pmb \chi}^{(2)}_{-{\bf q}}
    \end{pmatrix}\nonumber\\
    &\equiv -i\pi {\cal M}_{\alpha \beta}({\bf k},{\bf k'}),
\end{align}
with which Eq.~(\ref{dNdt}) becomes
\begin{align}
    \partial_t {\rho}_{\bf q}|_{\rm inj}&=\frac{\pi}{\hbar}\frac{S}{2N_{\rm int}}\frac{a^2}{d^2}\sum_{{\bf k},{\bf k'}}\sum_{\alpha,\beta}[f_{{\bf k'},\alpha}(1-f_{{\bf k},\beta})(\sigma_3+{\rho}_{\bf q})\nonumber\\
    &-f_{{\bf k},\beta}(1-f_{{\bf k'},\alpha})\rho_{\bf q}]{\cal M}_{\alpha \beta}({\bf k},{\bf k}')\delta_{{\bf k}+{\bf q},{\bf k}'}+{\rm H.c.}\nonumber\\
    &=-\frac{\pi}{\hbar}\frac{S}{2N_{\rm int}}\frac{a^2}{d^2}\sum_{{\bf k},{\bf k}'}
    \sum_{\alpha \beta}\left[{\rho}_{\bf q}-\sigma_3 N_0(\varepsilon_{{\bf k}',\alpha}-\varepsilon_{{\bf k},\beta})\right]\nonumber\\
    &\times(f_{{\bf k},\beta}-f_{{\bf k}',\alpha}){\cal M}_{\alpha \beta}({\bf k},{\bf k}')\delta_{{\bf k}+{\bf q},{\bf k}'}+{\rm H.c.}, \nonumber\\
    &=\frac{1}{\hbar}\frac{Sa^2}{2d^2}\sum_{\alpha \beta}(\rho_{\bf q}{\cal F}_{\alpha \beta}^{(1)}-\sigma_3{\cal F}_{\alpha \beta}^{(2)}),
    \label{dNdt_final}
\end{align}
in which we use $f_{{\bf k}',\alpha}(1-f_{{\bf k},\beta})=N_0(\varepsilon_{{\bf k}',\alpha}-\varepsilon_{{\bf k},\beta})(f_{{\bf k},\beta}-f_{{\bf k}',\alpha})$, where $N_0(\varepsilon)=\{\exp[{\varepsilon}/(k_BT)]-1\}^{-1}$ is the Bose-Einstein distribution at temperature $T$, ${\cal F}_{\alpha \beta}^{(1)}=-(\pi/N_{\rm int})\sum_{{\bf k},{\bf k}'}(f_{{\bf k},\beta}-f_{{\bf k}',\alpha}){\cal M}_{\alpha \beta}({\bf k},{\bf k}')\delta_{{\bf k}+{\bf q},{\bf k}'}$, and ${\cal F}_{\alpha \beta}^{(2)}=-(\pi/N_{\rm int})\sum_{{\bf k},{\bf k}'}N_0(\varepsilon_{{\bf k}',\alpha}-\varepsilon_{{\bf k},\beta})(f_{{\bf k},\beta}-f_{{\bf k}',\alpha}){\cal M}_{\alpha \beta}({\bf k},{\bf k}')\delta_{{\bf k}+{\bf q},{\bf k}'}$.
The detailed balance is checked by $\partial_t \langle \hat{\rho}_{\bf q}\rangle|_{\rm inj}=0$ at thermal equilibrium.

The coherent dynamics are given by
\begin{align}
    \partial_t {\rho}_{\bf q}|_{\rm coh}=\frac{1}{i\hbar}\left\langle\left[\hat{\rho}_{\bf q}^I(t),\hat{H}_m\right]\right\rangle =\frac{1}{i\hbar}\left({\rho}_{\bf q} \sigma_3{\cal E}_{\bf q}-{\cal E}_{\bf q}\sigma_3{\rho}_{\bf q}\right),
    \label{coherent_term}
\end{align}
which originates from the intrinsic spin torque Eq.~(\ref{torque}). Finally, using the relaxation-time approximation, we invoke the effect of the relaxation of magnons to its equilibrium density matrix $\rho_{\bf q}^{(0)}$ by including 
\begin{align}
    \partial_t {\rho}_{\bf q}|_{\rm rex}=-(\rho_{\bf q}-\rho_{\bf q}^{(0)})/\tau_m,
    \label{relaxation_term}
\end{align}
where $\tau_m$ is the relaxation time.

Contributed by Eqs.~\eqref{dNdt_final}, \eqref{coherent_term}, and \eqref{relaxation_term}, the quantum kinetic equation read
\begin{align}
    \partial_t {\rho}_{\bf q}=\partial_t {\rho}_{\bf q}|_{\rm inj}+\partial_t {\rho}_{\bf q}|_{\rm coh}+\partial_t {\rho}_{\bf q}|_{\rm rex}.
\end{align}

\section{Linear response regime}

\label{linear_response_regime}

For the spin injection of magnons by electrons, we consider the electron spin polarized along the N\'eel-vector $\hat{\bf z}$-direction with spin accumulation $\delta\mu=\delta\mu_{\uparrow}-\delta\mu_{\downarrow}$ in the heavy metal. 
The spin accumulation $\delta\mu_{\uparrow/\downarrow}\ll \varepsilon_{\rm F}$, so the population of electrons $f(\varepsilon_{{\bf k},\beta})=f_0(\varepsilon_{{\bf k}}-\varepsilon_{F})-\delta \mu_{\beta}{\partial f_0}/{\partial \varepsilon_{{\bf k}}}$, in which
$\varepsilon_{{\bf k}}=\hbar^2 k^2/(2m)$ with $m$ being the mass of electrons, $\varepsilon_{F}$ is the Fermi energy, and $f_0(\varepsilon_{{\bf k}}-\varepsilon_{F})=\{{\rm exp}[(\varepsilon_{{\bf k}}-\varepsilon_{F})/(k_BT)]+1\}^{-1}$ is the equilibrium Fermi-Dirac distribution of electrons. At room temperature, $k_BT\ll \varepsilon_F$, such that ${\partial f_0}/{\partial \varepsilon_{{\bf k}}}=-\delta(\varepsilon_{{\bf k}}-\varepsilon_{F})$.

We substitute the electron steady-state distribution into Eq.~(\ref{dNdt_final}) to calculate the spin injection to magnons. 
For the term ${\cal F}_{\alpha \beta}^{(1)}$ in Eq.~(\ref{dNdt_final}), we calculate
\begin{align}
    &-\frac{\pi}{N_{\rm int}}\sum_{{\bf k},{\bf k'}}(f_{{\bf k},\beta}-f_{{\bf k'},\alpha})\delta(\varepsilon_{{\bf k}}-\varepsilon_{{\bf k'}}+ \hbar\omega_{\zeta,{ \bf q}})\delta_{{\bf k}+{\bf q},{\bf k'}}\nonumber\\   
    &=C_q(\delta\mu_{\alpha}-\delta\mu_{\beta}-\hbar\omega_{\zeta,{\bf q}})
    \Theta\left[1-\left(\frac{m\omega_{\zeta,{\bf q}}}{\hbar k_{\rm F}q}-\frac{q}{2k_{\rm F}}\right)^2\right],
    \label{sumk1}
\end{align}
where $V=N_{\rm int}a^2d$ is the volume of the normal metal film and $C_q=a^2dm^2/(4\pi\hbar^4q)$.
The Heavyside step function $\Theta(x)$ indicates that the magnon excitation occurs only when $q$ exceeds a certain critical value. 
For small $q$, the magnon energy varies with $q^2$, while the electron band energy difference $\varepsilon_{{\bf k}+\bf q}-\varepsilon_{{\bf k}}$ approximately scales linearly with $q$.
Therefore, when $q \rightarrow 0$, the magnon energy is insufficient to match the energy variation of the electron band, leading to the prohibition of scattering. Similarly, for the magnon absorption
\begin{align}
    &-\frac{\pi}{N_{\rm int}}\sum_{{\bf k}}(f_{{\bf k},\beta}-f_{{\bf k}+{\bf q},\alpha})\delta(\varepsilon_{{\bf k}}-\varepsilon_{{\bf k}+\bf q}-\hbar\omega_{\zeta,-{\bf q}})\nonumber\\
    &=C_q(\delta\mu_{\alpha}-\delta\mu_{\beta}+\hbar\omega_{\zeta,-{\bf q}})
    \Theta\left[1-\left(\frac{m\omega_{\zeta,-{\bf q}}}{\hbar k_{\rm F}q}+\frac{q}{2k_{\rm F}}\right)^2\right].
    \label{sumk2}
\end{align}
Since $q$ is much smaller than the Fermi wave vector $k_F \approx 10^{10}$~m$^{-1}$, $q/(2k_F)\ll 1$ such that 
\[
\Theta_{\zeta,\pm {\bf q}}\equiv \Theta\left[1-\left(\frac{m\omega_{\zeta,\pm{\bf q}}}{\hbar k_{\rm F}q}\mp\frac{q}{2k_{\rm F}}\right)^2\right]\approx \Theta\left[1-\frac{m\omega_{\zeta,\pm{\bf q}}}{\hbar k_{\rm F}q}\right].
\]
In the second term ${\cal F}_{\alpha \beta}^{(2)}$ of Eq.~(\ref{dNdt_final}), 
\begin{align}
    &-\frac{\pi}{N_{\rm int}}\sum_{{\bf k}}N_0(\varepsilon_{{\bf k}+{\bf q},\alpha}-\varepsilon_{{\bf k},\beta})\nonumber\\
    &\times(f_{{\bf k},\beta}-f_{{\bf k}+{\bf q},\alpha})\delta(\varepsilon_{{\bf k}}-\varepsilon_{{\bf k}+{\bf q}}\pm \hbar\omega_{\zeta,\pm{\bf q}}) \nonumber\\
    &=N_0(\pm \hbar\omega_{\zeta,\pm {\bf q}}-\delta\mu_{\alpha}+\delta\mu_{\beta})\nonumber\\
    &\times
    \left[-\frac{\pi}{N_{\rm int}}\sum_{\bf k}(f_{{\bf k},\beta}-f_{{\bf k}+{\bf q},\alpha})\delta(\varepsilon_{{\bf k}}-\varepsilon_{{\bf k}+{\bf q}}\pm \hbar\omega_{\zeta,\pm{\bf q}})\right].
    \label{sumk3}
\end{align}

According to Eqs.~(\ref{sumk1}), (\ref{sumk2}), and (\ref{sumk3}), Eq.~(\ref{dNdt_final}) can be rewritten as
\begin{align}
     \partial_t {\rho}_{\bf q}|_{\rm inj}=\frac{1}{\hbar}\frac{Sa^2}{2d^2}C_q\sum_{\alpha \beta}\big[\rho_{\bf q}-\sigma_3 {\cal N}_{\alpha\beta}({\bf q})\big]{\cal R}_{\alpha \beta}({\bf q})+{\rm H.c.},
     \label{inj_final}
\end{align}
in which the matrices 
\begin{align}
    {\cal N}_{\alpha\beta}({\bf q})&={\rm diag}\{N_0(\hbar\omega_{1,{\bf q}}-\delta\mu_{\alpha\beta}),N_0(\hbar\omega_{2,{\bf q}}-\delta\mu_{\alpha\beta}),\nonumber\\
    &N_0(-\hbar\omega_{1,-{\bf q}}-\delta\mu_{\alpha\beta}),N_0(-\hbar\omega_{2,-{\bf q}}-\delta\mu_{\alpha\beta})\}\nonumber
\end{align}
with $\delta\mu_{\alpha\beta}=\delta\mu_{\alpha}-\delta\mu_{\beta}$,
and  
\begin{align}
    &{\cal R}_{\alpha \beta}({\bf q})=\begin{pmatrix}
    (\delta\mu_{\alpha}-\delta\mu_{\beta}-\hbar\omega_{1,{\bf q}})\Theta_{1,{\bf q}}{\pmb \sigma}_{\alpha \beta}\cdot {\pmb \chi}^{(1)}_{{\bf q}} \\
    (\delta\mu_{\alpha}-\delta\mu_{\beta}-\hbar\omega_{2,{\bf q}})\Theta_{2,{\bf q}}{\pmb \sigma}_{\alpha \beta}\cdot {\pmb \chi}^{(2)}_{{\bf q}} \\
    (\delta\mu_{\alpha}-\delta\mu_{\beta}+\hbar\omega_{1,-{\bf q}})\Theta_{1,-{\bf q}}{\pmb \sigma}_{\alpha \beta}\cdot {\pmb \chi}_{-{\bf q}}^{(1)\ast} \\
    (\delta\mu_{\alpha}-\delta\mu_{\beta}+\hbar\omega_{2,-{\bf q}})\Theta_{2,-{\bf q}}{\pmb \sigma}_{\alpha \beta}\cdot {\pmb \chi}_{-{\bf q}}^{(2)\ast}
    \end{pmatrix} \nonumber\\
    &\times
    \begin{pmatrix}
        {\pmb \sigma}_{\beta \alpha}\cdot {\pmb \chi}_{{\bf q}}^{(1)\ast}, & 
        {\pmb \sigma}_{\beta \alpha}\cdot {\pmb \chi}_{{\bf q}}^{(2)\ast}, &
        -{\pmb \sigma}_{\beta \alpha}\cdot {\pmb \chi}^{(1)}_{-{\bf q}}, &
        -{\pmb \sigma}_{\beta \alpha}\cdot {\pmb \chi}^{(2)}_{-{\bf q}}
    \end{pmatrix}.
\end{align}
According to Eq.~(\ref{inj_final}), the detailed balance is guaranteed with  $\partial_t \langle \hat{\rho}_{\bf q}\rangle|_{\rm inj}=0$ in the absence of the spin accumulation  $\delta\mu_{\alpha}=\delta\mu_{\beta}=0$.

\end{appendix}

\end{document}